\documentclass[submission,copyright,creativecommons]{eptcs} 
\providecommand{\event}{FVAV 2017}
\usepackage{etex} 
\usepackage{graphicx, tikz}
\usepackage{times}

\usepackage{breakurl}
\usepackage{underscore}

\usepackage{theorem}
\usepackage{fancybox,amssymb,amstext,amsmath,stmaryrd,wasysym,cite,proof,mathtools,multirow,stackrel,subfigure,here,hhline}
\SetSymbolFont{stmry}{bold}{U}{stmry}{m}{n}
\SetSymbolFont{wasy}{bold}{U}{wasy}{m}{n}

\usetikzlibrary{shapes,arrows,positioning}
\usetikzlibrary{decorations.text}

\usepackage{algorithm}
\usepackage{algpseudocode}
\algnewcommand\algorithmicinput{\textbf{Input:}}
\algnewcommand\INPUT{\item[\algorithmicinput]}
\algnewcommand\algorithmicannotation{\textbf{Annotation:}}
\algnewcommand\ANNOTATION{\item[\algorithmicannotation]}
\algnewcommand\algorithmicoutput{\textbf{Output:}}
\algnewcommand\OUTPUT{\item[\algorithmicoutput]}
\algnewcommand\algorithmichyperparam{\textbf{Hyper parameters:}}
\algnewcommand\HYPERPARAM{\item[\algorithmichyperparam]}
\makeatletter
\algrenewcommand\ALG@beginalgorithmic{\footnotesize}
\makeatother
\allowdisplaybreaks[1] 

\newif\ifignore 
\ignorefalse
\newcommand{\auxproof}[1]{
  \ifignore\mbox{}\newline
  \textbf{BEGIN: AUX-PROOF} \dotfill\newline
  {#1}\mbox{}\newline
  \textbf{END: AUX-PROOF}\dotfill\newline
  \fi}

\usepackage{wrapfig}
\theorembodyfont{\itshape}
\newtheorem{mytheorem}{Theorem}[section]

\theorembodyfont{\rmfamily}
\newtheorem{mynotation}[mytheorem]{Notation}
\newtheorem{myremark}[mytheorem]{Remark}
\newtheorem{myexample}{Example}
\newtheorem{mydefinition}[mytheorem]{Definition}

\newcommand{\R}{{\mathbb{R}}}

\newcommand{\Models}{{ \;\mathbin{\vDash}\; }}
\newcommand{\NModels}{{ \;\mathbin{\nvDash}\; }}

\newcommand{\Defiff}{\stackrel{\text{def.}}{\Longleftrightarrow}}

\newcommand{\ttrue}{\mathrm{t{\kern-1.5pt}t}}
\newcommand{\ffalse}{\mathrm{f{\kern-1.5pt}f}}

\newcommand{\place}{\underline{\phantom{n}}\,}

\newcommand{\STL}{\textbf{STL}}
\newcommand{\MITL}{\textbf{MITL}}
\newcommand{\GPUCB}{\textbf{GP-UCB}}
\newcommand{\GPPS}{\textbf{GP-PSat}}
\newcommand{\GPPI}{\textbf{GP-PI}}

\newcommand{\KLDist}{D_{\textsf{KL}}}

\newcommand{\UntilOp}[1]{\mathbin{\mathcal{U}_{#1}}}

\newcommand{\DiaOp}[1]{\Diamond_{#1}}
\newcommand{\BoxOp}[1]{\square_{#1}}

\newcommand{\Robust}[2]{{ \llbracket #1,\, #2 \rrbracket}}
\newcommand{\sem}[1]{\llbracket #1 \rrbracket} 
\newcommand{\Defeq}{\triangleq}

\newcommand{\Vee}[1]{{{\bigsqcup_{#1}}}}
\newcommand{\Wedge}[1]{{{\bigsqcap_{#1}}}}

\newcommand{\Argmin}{\operatornamewithlimits{\mathrm{arg\,min}}}

\newcommand{\GP}{\mathbf{GP}}
\newcommand{\Prob}{\mathrm{Pr}}

\newcommand{\Rnn}{\R_{\ge 0}}

\usepackage{enumitem}

\title{
  Causality-Aided
  Falsification
}
\author{
  Takumi Akazaki
  \thanks{  Supported by Grants-in-Aid for JSPS Fellows
    No.\ 15J09877.}
  \institute{
    The University of Tokyo, Japan\\
    JSPS Research Fellow
  }
  \email{akazaki@ms.k.u-tokyo.ac.jp}
  \and
  Yoshihiro Kumazawa
  \institute{
    The University of Tokyo, Japan
  }
  \email{kumazawa@is.s.u-tokyo.ac.jp}
  \and
  Ichiro Hasuo
  \thanks{
    Supported by JST ERATO HASUO Metamathematics for Systems
    Design Project (No. JPMJER1603), and JSPS Grant-in-Aid No. 15KT0012.
  }
  \institute{
    National Institute of Informatics, Tokyo, Japan
  }
  \email{hasuo@nii.ac.jp}
}
\def\titlerunning{Causality-Aided Falsification}
\def\authorrunning{T. Akazaki, Y. Kumazawa, and I. Hasuo}
\begin{document}

\maketitle

\begin{abstract}
 \emph{Falsification} is drawing attention in quality assurance of heterogeneous  systems whose complexities are beyond most verification techniques' scalability. In this paper we introduce the idea of \emph{causality aid} in falsification: by providing a falsification solver---that relies on stochastic optimization of a certain cost function---with suitable causal information expressed by a \emph{Bayesian network}, search for a falsifying input value can be efficient. Our experiment results show the idea's viability. 
\end{abstract}

\section{Introduction}\label{sec:Introduction}
\textbf{Falsification}\quad
In computer science, \emph{verification} refers to the task of giving a mathematical proof to the claim that the behavior of a system $\mathcal{M}$ satisfies a  desired property $\varphi$ (called a \emph{specification}), under any circumstances (such as  choices of input to the system $\mathcal{M}$).  A mathematical proof thus obtained gives a level of confidence that is fundamentally different from empirical guarantees given by \emph{testing}. 

Extensive research efforts have yielded a number of effective verification techniques and they have seen successful real-world applications. 
At the same time, however, it is also recognized that \emph{large-scale heterogeneous systems} are still beyond the scalability of most of these verification techniques. 
Notable among such are \emph{cyber-physical systems (CPSs)} that exhibit not only discrete digital dynamics but also continuous physical dynamics. Imagine a car today: it  contains not only dozens of chips (like ECUs) but also continuous dynamics (wheels, suspensions, internal combustion, etc.).

It is in this CPS context that the idea of \emph{falsification} is found its use~\cite{DBLP:conf/cpsweek/HoxhaAF14a}.
 \begin{quote}
    \underline{\bfseries The falsification problem}
    \begin{itemize}
    \item{\textbf{Given:}} 
      a \emph{model} $\mathcal{M}$ (a function from an input signal
      to  an output signal),
      and
      a \emph{specification} $\varphi$ (a temporal formula)
    \item{\textbf{Answer:}} 
      a \emph{critical path}, that is, an input signal $\sigma_{\mathrm{in}}$ such
      that the corresponding output $\mathcal{M}(\sigma_{\mathrm{in}})$ does not
      satisfy $\varphi$ 
    \end{itemize}
\end{quote}
 Two benefits of falsification are particularly appealing. For one, a system model $\mathcal{M}$ can be totally a \emph{black box}: once $\mathcal{M}$ as a function $\sigma_{\mathrm{in}}\mapsto \mathcal{M}(\sigma_{\mathrm{in}})$ is given as an oracle, we can check if an educated guess $\sigma_{\mathrm{in}}$ is  a solution or not---without knowing $\mathcal{M}$'s internal working.
This is an advantage given that many CPSs do have black-box components:
they can come from external suppliers, or they can be physical dynamics too complex to mathematically model (typically one uses \emph{look-up tables} to describe such). 

Another appealing feature of falsification is its affinity with \emph{machine learning (ML)} and \emph{optimization} techniques. In automatic verification techniques the greatest challenge is \emph{state-space explosion}: the size of the  input space for $\sigma_{\mathrm{in}}$ grows exponentially with respect to its dimension, often to the extent that exhaustive search in it is no longer possible. Recent surges in ML and optimization algorithms can offer potent countermeasures against this \emph{curse of dimensionality}: specifically, after observing
output
$\mathcal{M}(\sigma_{1}),\dotsc,\mathcal{M}(\sigma_{n})$ for 
 input $\sigma_{1},\dotsc,\sigma_{n}$, those algorithms can ``learn'' from these previous attempts and  suggest an input signal 
$\sigma_{n+1}$ with which $\mathcal{M}(\sigma_{n+1})\not\models \varphi$ is likely.

One can say that falsification is after all \emph{adaptive testing}: most falsification solvers rely on stochastic guess; hence their failure do not prove ``$\mathcal{M}(\sigma_{\mathrm{in}})\models\varphi$ for every $\sigma_{\mathrm{in}}$.'' However in many real-world scenarios falsification is as good as one gets, because of systems' complexity and their black-box components within. Existing stochastic optimization-based solvers (such as
S-TaLiRo~\cite{DBLP:conf/tacas/AnnpureddyLFS11}
and
BREACH~\cite{DBLP:conf/cav/Donze10}) have shown striking performance, too, scaling up to various Simulink diagrams from automotive applications. Moreover, falsification has special appeal to real-world engineers: while it takes certain familiarity  to come to appreciate correctness proofs,  counterexamples discovered by falsification  easily convince engineers that there are issues to be resolved. 
 
\noindent
\textbf{Search of Cost Functions}\quad
A technical cornerstone that set off the study of falsification is \emph{robust semantics} of temporal formulas~\cite{DBLP:conf/formats/DonzeM10, DBLP:journals/tcs/FainekosP09}. 
With CPS application in mind we assume that input and output of our system model $\mathcal{M}$ are given by (time-variant) signals. For them it is standard to specify 
 properties  using some \emph{temporal logic}, such as 
metric interval temporal logic 
($\textbf{MITL}$)~\cite{DBLP:journals/jacm/AlurFH96}
and 
signal temporal logic 
($\textbf{STL}$)~\cite{DBLP:conf/formats/MalerN04}.
In robust semantics~\cite{DBLP:conf/formats/DonzeM10, DBLP:journals/tcs/FainekosP09} a signal $\sigma$ and a
formula $\varphi$ are assigned a \emph{continuous} truth value $\Robust{\sigma}{\varphi}\in\R$
that designates how robustly the formula is satisfied. This departure from the conventional Boolean semantics (where $\Robust{\sigma}{\varphi}\in\{\ttrue,\ffalse\}$) allows one to adopt a hill climbing-style optimization algorithm to look for a falsifying input signal. 

Algorithm~\ref{alg:falsificationByOptHighLevel} is 
a   high-level description of falsification by optimization. 
Here a \emph{cost function} $f_{\varphi}$ carries a signal (output of the system $\mathcal{M}$) to a real; we assume that its value is linked with satisfaction of $\varphi$, that is specifically,  $f_{\varphi}(\tau)<0$ implies $\tau\not\models\varphi$. We assume that the value of $f_{\varphi}$ for a given input can be effectively computed; we assume the same for the function $\mathcal{M}$. Still in Line~\ref{Line:algfalsificationByOptHighLevelGuess} the true solution may not be available since the global structure of $\mathcal{M}$ is  unknown---this reflects 
our black-box view on   $\mathcal{M}$. Therefore in Line~\ref{Line:algfalsificationByOptHighLevelGuess} we make a guess based on the previous trials. 

\begin{algorithm}[tbp]
  \caption{Falsification by optimization, with a cost function $f_{\varphi}$}
  \label{alg:falsificationByOptHighLevel}
  \begin{algorithmic}[1]
    \INPUT  $\sigma_{0}$ \Comment The initial guess
    \State $v_{0}:=f_{\varphi}(\mathcal{M}(\sigma_{0}))$;
    \Comment $v_{i}=f_{\varphi}(\mathcal{M}(\sigma_{i}))$ is the score of 
    the input  $v_{i}$
    \For{ $i = 1 \dots N$ }
    \Comment $N$ is the greatest number of iteration
    \State\label{Line:algfalsificationByOptHighLevelGuess}
    $\sigma_{i}:=\Argmin_{\sigma}
    \left(\;
      f_{\varphi}(\mathcal{M}(\sigma))
      \;\bigg|\quad
      \begin{minipage}{.4\textwidth}
	under the previous observations
	
        $(\sigma_{0},v_{0}),
        (\sigma_{1},v_{1}),
        \dotsc,
        (\sigma_{i-1},v_{i-1})
        $
      \end{minipage}
      \;
    \right)
    $;
    \State $v_{i}:=f_{\varphi}(\mathcal{M}(\sigma_{i}))$;
    \If{$v_{i}<0$}
    \Return $\sigma_{i}$;
    \EndIf
    \Comment{Falsification succeeded, because we assume
      $f_{\varphi}(\tau)<0$ implies $\tau\not\models\varphi$}
    \State $i:=i+1$;
    \EndFor
  \end{algorithmic}
\end{algorithm}

The robust semantics of temporal formulas in~\cite{DBLP:conf/formats/DonzeM10, DBLP:journals/tcs/FainekosP09} is a prototype of such a cost function (Algorithm~\ref{alg:falsificationByOptHighLevel}). 
 Subsequently in the study of falsification,
search of better cost functions has been an important topic. For example, sometimes \emph{time robustness}~\cite{DBLP:conf/formats/DonzeM10}---as opposed to \emph{space robustness} in the original work~\cite{DBLP:journals/tcs/FainekosP09}---yields smoother hills to climb down, aiding optimization. Combination of space and time robustness is pursued in~\cite{DBLP:conf/cav/AkazakiH15}, where they enrich logics with \emph{averaged} modalities to systematically enhance expressivity. Additional bias is put on cost functions in~\cite{DBLP:conf/emsoft/Dokhanchi0SSF15} so that search for falsifying input covers a greater number of discrete modes of a system $\mathcal{M}$.  After all, the question here is how to enrich cost functions, extracting additional information from a system $\mathcal{M}$ and/or a specification $\varphi$.

\vspace{.5em}
\noindent
\textbf{Contribution: Causality Aid in Falsification}\quad
In this paper we build on the  observations in~\cite{DBLP:conf/rv/Akazaki16} and propose to aid falsification using \emph{causal information}. We lay out the idea using a simple example. 

\vspace{.5em}
\noindent
\begin{minipage}{.7\textwidth}
 \begin{myexample}[incremental counter]\label{ex:counter}
 Consider the pseudocode shown on the right.
 We think of: $i_0, i_1, \dots i_N \in [-1,1]$  as the values of a time-variant input signal  $i$ at time $t=0,1,\dotsc, N$, respectively;  Lines~\ref{line:counterFirst}--\ref{line:counterLast} as a system $\mathcal{M}$ that takes such input and returns the value of $\mathsf{cnt}$ as output; and the assertion $\mathsf{cnt} \le N$ as a specification $\varphi$, that is $\Box_{[N,N]}(\mathsf{cnt}\le N)$ in temporal logic. It is clear that, to falsify $\varphi$, all the input values $i_0, i_1, \dotsc, i_N$ must lie in $(-0.2,0.2)$; otherwise the counter is reset and never reaches $\mathsf{cnt}=N+1$. 
 \end{myexample}
\end{minipage}
\quad
\scalebox{.82}{\begin{minipage}{15em}
        \begin{algorithmic}[1]
         \INPUT $i_0, i_1, \dotsc, i_N \in [-1,1]$
         \OUTPUT $\mathsf{cnt}$
         \State \label{line:counterFirst} $t:=0; \mathsf{cnt} :=0$
         \While{$t \leq N$}
         \State $\mathsf{flag} := (|i_t| < 0.2)$;
         \If{$\mathsf{flag}$}
         \State $\mathsf{cnt} := \mathsf{cnt} + 1$
         \Else
         \State $\mathsf{cnt} := 0$;
         \EndIf
         \State $t := t+1$;
         \EndWhile \label{line:counterLast}
         \State $\textbf{assert}(\mathsf{cnt} \le N)$
       \end{algorithmic}
\end{minipage}
}

Now consider solving the falsification problem here. It turns out that
existing falsification solvers have hard time in doing so: besides the
apparent hardness of the problem (following the uniform distributions the
success  probability would be $0.2^{N}$), there is the following ``causal''
reason for the difficulty.

Assume $i_{N}\not\in (-0.2,0.2)$, meaning that
$\mathsf{cnt}$ is reset to $0$ at the last moment. In this case
the earlier input values $i_0, i_1, \dotsc, i_{N-1}$ have no effect in
the final output $\mathsf{cnt}$, nor in the robust semantics of the
specification $\Box_{[N,N]}(\mathsf{cnt}\le N)$. Therefore there is no
incentive for stochastic optimization solvers to choose values
$i_0, \dotsc, i_{N-1}$ from $(-0.2,0.2)$. More generally, 
desired bias is imposed on earlier input values
$i_0, \dotsc, i_{k}$ \emph{only after}
later input values    $i_{k+1}, \dotsc, i_{N}$ have been suitably
fixed. Given the system (the above program) as a black box and the specification
$\Box_{[N,N]}(\mathsf{cnt}\le N)$ alone, there is no way for
optimization solvers to know such causal dependency.

\begin{wrapfigure}[13]{R}{0.4\textwidth}
  \centering
  \scalebox{.8}{
    \begin{minipage}{.4\textwidth}
      \begin{tikzpicture}[
        node distance=1cm and 0cm,
        mynode/.style={draw,ellipse,text width=2.45cm,align=center}
        ]
        \node[mynode] (phi5) 
        {$\varphi_5= \BoxOp{[5,5]}(\mathsf{cnt} \leq 5)$};
        \node[mynode,above=0.3cm of phi5] (phi4)
        {$\varphi_4= \BoxOp{[4,4]}(\mathsf{cnt} \leq 4)$};
        \coordinate[above=0.3cm of phi4] (p4);
        \coordinate[above=0.3cm of p4] (p1);
        \node[mynode,above=0.3cm of p1] (phi0)
        {$\varphi_0= \BoxOp{[0,0]}(\mathsf{cnt} \leq 0)$};
        \node[rectangle callout,draw,inner sep=2pt,
        callout absolute pointer=(phi0.west),
        left= 0.5cm of phi0](b0){
          \begin{minipage}{.44\textwidth}
            \centering
            \begin{tabular}{ p{0.4cm}||cc }
              \qquad
                & $ \ttrue$
                & $ \ffalse$
              \\ \hhline{=||==}
                & 0.8 & 0.2 \\
            \end{tabular}
          \end{minipage}};
        \node[rectangle callout,draw,inner sep=2pt,
        callout absolute pointer=(phi4.west),
        above left= -0.59cm and 1.07cm of phi4](b4){
          \begin{minipage}{.44\textwidth}
            \centering
            \begin{tabular}{p{0.4cm}||cc }
              $\varphi_3$ 
                & $ \ttrue$
                & $ \ffalse$
              \\ \hhline{=||==}
              $\ttrue$ & 1 & 0 \\ 
              $\ffalse$ & 0.8 & 0.2 \\
            \end{tabular}
          \end{minipage}};
        \node[rectangle callout,draw,inner sep=2pt,
        callout absolute pointer=(phi5.west),
        left= 0.5cm of phi5](b5){
          \begin{minipage}{.44\textwidth}
            \centering
            \begin{tabular}{p{0.4cm}||cc }
              $\varphi_4$ 
                & $ \ttrue$
                & $ \ffalse$
              \\ \hhline{=||==}
              $\ttrue$ & 1 & 0 \\ 
              $\ffalse$ & 0.8 & 0.2 \\
            \end{tabular}
          \end{minipage}};
        \path
        (phi4) edge[-latex] (phi5)
        (p4) edge[-latex] (phi4)
        (phi0) edge[-latex] (p1);
        \draw[dotted] (p1) -- (p4);
      \end{tikzpicture}
    \end{minipage}}
  \caption{Causality annotation for incremental counter (where $N=5$)}
  \label{fig:BNExample1Intro}
\end{wrapfigure}
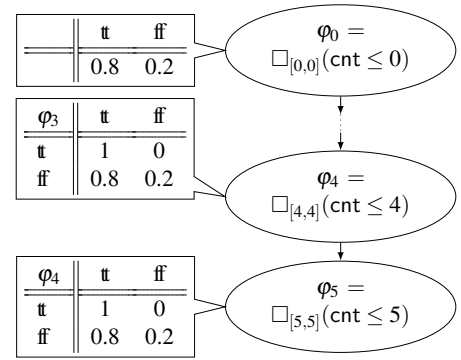

Our enhancement of falsification algorithms consists of leveraging
causal information expressed as \emph{Bayesian networks}. See
Fig.~\ref{fig:BNExample1Intro}, where we fix $N=5$ for presentation.  The Bayesian
network expresses causal
dependence of the original specification $\varphi=\varphi_{5}$ on
other specifications
$\varphi_{0},\dotsc,\varphi_{4}$. The newly introduced specifications
$\varphi_{i}= \BoxOp{[i,i]}(\mathsf{cnt} \leq i)$, for each $i=0,\dotsc,4$, express that the counter $\mathsf{cnt}$ has already been reset by time $i$. Therefore in order to falsify $\varphi_{5}$, i.e.\ to keep incrementing  $\mathsf{cnt}$, these additional specifications must be falsified, too. The last observation is expressed in the Bayesian network in Fig.~\ref{fig:BNExample1Intro}, specifically in the conditional probabilities $\Prob\bigl(\sem{\varphi_{i+1}}=\ffalse\mid \sem{\varphi_{i}}=\ttrue\bigr)=0$. 

Now our falsification algorithm looks not only at $\varphi_{5}$ but also at the other predicates $\varphi_{0},\dotsc,\varphi_{4}$. This way we successfully impose bias on earlier input values $i_0, \dotsc, i_{4}$ to lie in $(-0.2,0.2)$---as demonstrated by our experimental results later. 

\vspace{.5em}
Following the idea illustrated in the last example, our main contribution in this paper is a \emph{causality-aided} falsification algorithm that uses \emph{Bayesian networks of temporal formulas} as input on the specification side. Such a Bayesian network can be derived from an original specification $\varphi$ alone; they can also be derived through inspection of a system model $\mathcal{M}$; or we can use both $\varphi$ and $\mathcal{M}$. In order to efficiently leverage the causal information expressed by a Bayesian network, we follow~\cite{DBLP:journals/corr/ChenSK16a,DBLP:conf/cdc/ChenSK16} and use variations of \emph{Gaussian process optimization} as our optimization algorithms (Line~\ref{Line:algfalsificationByOptHighLevelGuess} of Algorithm~\ref{alg:falsificationByOptHighLevel}). The feature that they allow to guess both average and variance (see~\S{}\ref{subsec:GPOptimization}) turns out to be particularly useful. 
We implemented the algorithm; our experimental results, although preliminary, seem to support the effectiveness of our approach.

General methodologies of deriving such Bayesian networks are outside the paper's focus, although we do have some preliminary ideas and we  exploited them for our current examples. One is the use of \emph{probabilistic predicate transformers} that are a classic topic in semantics~\cite{Kozen81,Jones90PhD,MorganMS96} and are shed fresh light on in the context of  \emph{probabilistic programming languages} (see e.g.~\cite{DBLP:conf/lics/OlmedoKKM16}). This idea follows the earlier observations in~\cite{DBLP:journals/corr/AkazakiHS15}; it successfully generates
 the Bayesian network in Fig.~\ref{fig:BNExample1Intro}
for Example~\ref{ex:counter}.
Another idea is parse tree-like decomposition of an original temporal formula $\varphi$; we decorate the resulting tree with conditional probabilities that we learn through sampling. These methods will be described in our  forthcoming papers.



\vspace{.5em}
\noindent
\textbf{Related Work}\quad
Besides search of better cost functions,
an important direction in the study of falsification is
improving  optimization algorithms (that are used in Line~\ref{Line:algfalsificationByOptHighLevelGuess} of Algorithm~\ref{alg:falsificationByOptHighLevel}). In the falsification literature many different algorithms have been used and studied: they include
\emph{simulated annealing}, \emph{ant-colony optimization},   the \emph{cross-entropy method},  the \emph{Nelder-Mead algorithm}, and so on~\cite{Sankaranarayanan:2012:FTP:2185632.2185653,DBLP:conf/cav/Donze10,DBLP:conf/tacas/AnnpureddyLFS11}
.
In~\cite{DBLP:conf/atva/DeshmukhJKM15} a discrete algorithm of \emph{Tabu search} is employed for enhanced coverage. 

Yet another important direction is \emph{multiple shooting falsification}~\cite{Zutshi:2014:MSC:2656045.2656061,DBLP:conf/cdc/ZutshiSDK13} where, unlike \emph{single shooting} approaches like in this paper, a bunch of trajectories are investigated in a single iteration relying on suitable abstraction of a system model and/or a specification.
We believe our idea of causality  aid in falsification is orthogonal to the choice between single and multiple shooting; we will study  as future work the effect of causality in multiple shooting falsification.

\section{Backgrounds}\label{sec:bck}

\subsection{STL and Robust Semantics}\label{subsec:STL}




Here we present \emph{signal temporal logic ($\STL$)}~\cite{DBLP:conf/formats/MalerN04} as our formalism for expressing (original, without causal information) specifications. 
We also present its \emph{robust semantics}~\cite{DBLP:conf/formats/DonzeM10} that give the prototype of the cost function $f_{\varphi}$ in Algorithm~\ref{alg:falsificationByOptHighLevel}. Our cost function will be derived from the robust semantics of the formulas in a Bayesian network. At the same time we emphasize that  our methodology of causality-aided falsification does not depend on the specific underlying specification formalism of $\STL$. 



\begin{mydefinition}[syntax of $\STL$]\label{sec:Syntax}
  The set of \emph{$\STL$ formulas} 
  are recursively defined as follows.
  \begin{align*}
    \varphi & \,::=\,
              g(\mathbf{y}) > 0 
              \mid \neg \varphi
              \mid \varphi_1 \vee \varphi_2
              \mid \varphi_1 \UntilOp{I} \varphi
  \end{align*}
 Here $g(\mathbf{y})$ is some real-value function
  over the set of variables $\mathbf{y} = \{y_1, \dots, y_n\}$,
  and $I$ is a closed non-singular interval in $\Rnn$.  
\end{mydefinition}
We also introduce
the following standard temporal operators as abbreviations:
the \emph{eventually} operator $\DiaOp{I}\varphi \Defeq (\infty > 0) \UntilOp{I} \varphi$ and
the  \emph{always} operator $\BoxOp{I}\varphi\Defeq \neg \DiaOp{I} \neg \varphi$.
\begin{mydefinition}[Boolean semantics of $\STL$]\label{def:BooleanSemantics}
  Let $\sigma_\mathbf{y}\colon   \Rnn\to \R^{n}$ be a \emph{signal}, that is, a function  
  that maps  time $\tau$
  to the values $\sigma_\mathbf{y}(\tau)$ of the variables $\mathbf{y}$ at time $\tau$.
  We define the (Boolean) validity of an $\STL$ formula over a signal $\sigma_\mathbf{y}$,
  as follows.
  Here $\sigma_\mathbf{y}^\tau$ stands for the time-shifted signal
  such that $\sigma_\mathbf{y}^\tau(\tau') \Defeq \sigma_\mathbf{y}(\tau + \tau')$.
  \begin{displaymath}
    \begin{array}{lrl}
      \sigma_\mathbf{y} \Models g(\mathbf{y}) > 0
      & \quad\Defiff\quad 
      & \text{the inequality } g(\sigma_\mathbf{y}(0)) > 0 \text{ holds}\\ 
      \sigma_\mathbf{y} \Models \neg \varphi
      & \quad\Defiff\quad 
      & \sigma_\mathbf{y} \NModels \varphi\\
      \sigma_\mathbf{y} \Models \varphi_1 \vee \varphi_2
      & \quad\Defiff\quad 
      & \sigma_\mathbf{y} \Models \varphi_1 
        \text { or } \sigma_\mathbf{y} \Models \varphi_2\\                 
      \sigma_\mathbf{y} \Models \varphi_1 \UntilOp{I} \varphi_2
      & \quad\Defiff\quad 
      & \exists \tau \in I.
        \big(
        \sigma_\mathbf{y}^\tau \Models \varphi_2 \text { and } 
        \forall \tau' \in [0,\tau]. \sigma_\mathbf{y}^{\tau'} \Models \varphi_1
        \big)\\                 
    \end{array}
  \end{displaymath}
\end{mydefinition}

The following ``quantitative refinement'' of the semantics of $\STL$ initiated the research program of falsification by optimization~\cite{DBLP:conf/formats/DonzeM10, DBLP:journals/tcs/FainekosP09}. 

\begin{mydefinition}[robust semantics of STL]\label{def:semantics}
  For a signal $\sigma_\mathbf{y}$ and an $\STL$ formula $\varphi$,
  we define the \emph{robustness} $\Robust{\sigma_\mathbf{y}}{\varphi} \in \R \cup\{\infty,-\infty\}$ 
  inductively as follows. 
  Here $\sqcap$ and $\sqcup$ denote infimums and supremums of real numbers, respectively.
  \begin{displaymath}
    \begin{array}{lrl}
      \Robust{\sigma_\mathbf{y}}{g(\mathbf{y}) > 0}
      & \quad\Defeq\quad
      & g(\sigma_\mathbf{y}(0))\\
      \Robust{\sigma_\mathbf{y}}{\neg \varphi}
      & \quad\Defeq\quad 
      & - \Robust{\sigma_\mathbf{y}}{\varphi}\\
      \Robust{\sigma_\mathbf{y}}{\varphi_1 \vee \varphi_2}
      & \quad\Defeq\quad 
      & \Robust{\sigma_\mathbf{y}}{\varphi_1}
        \sqcup \Robust{\sigma_\mathbf{y}}{\varphi_2}\\                 
      \Robust{\sigma_\mathbf{y}}{\varphi_1 \UntilOp{I} \varphi_2}
      & \quad\Defeq\quad
      & \Vee{\tau \in I}
        \big(
        \Robust{\sigma_\mathbf{y}^\tau}{\varphi_2} \sqcap
        \Wedge{\tau' \in [0,t]} \Robust{\sigma_\mathbf{y}^{\tau'}}{\varphi_1}
        \big)\\                       
    \end{array}
  \end{displaymath}
\end{mydefinition}

Note that 
the sign of robustness coincides with the Boolean semantics. 
That is,
$\Robust{\sigma_\mathbf{y}}{\varphi}>0$ implies $\sigma_\mathbf{y} \Models \varphi$, 
and
$\Robust{\sigma_\mathbf{y}}{\varphi}<0$ implies $\sigma_\mathbf{y} \NModels \varphi$.
Conversely,
$\sigma_\mathbf{y} \Models \varphi$ implies 
$\Robust{\sigma_\mathbf{y}}{\varphi}\geq 0$,
and
$\sigma_\mathbf{y} \NModels \varphi$ implies 
$\Robust{\sigma_\mathbf{y}}{\varphi}\leq 0$.


\subsection{Gaussian Process Optimization}\label{subsec:GPOptimization}

\begin{figure}[b]
  \centering
  \begin{minipage}{\textwidth}
    \centering
    \begin{minipage}{.3\textwidth}
      \centering{
      \includegraphics[width=\textwidth]{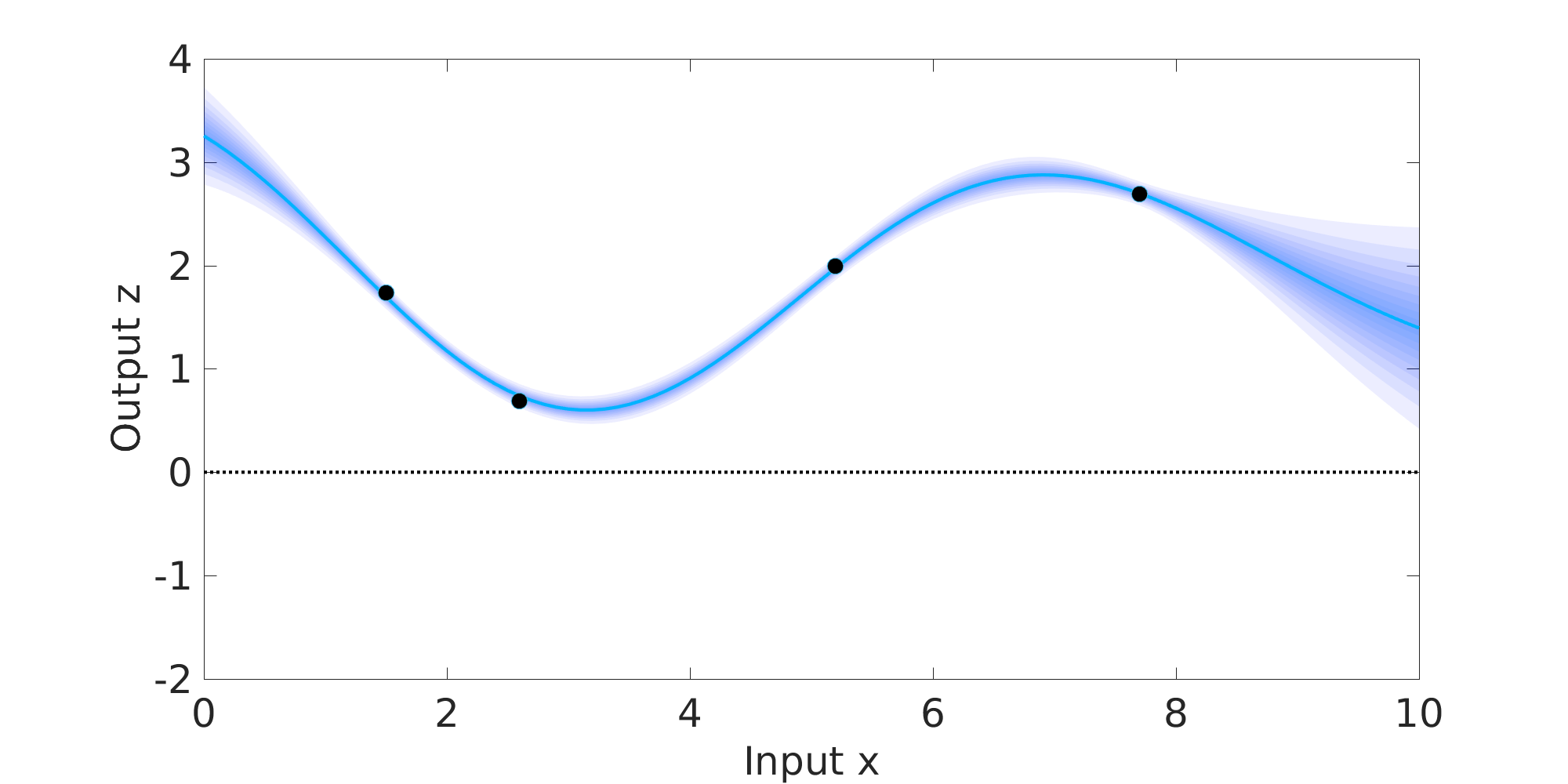}
      $l=1$}
    \end{minipage}
    \begin{minipage}{.3\textwidth}
      \centering{
      \includegraphics[width=\textwidth]{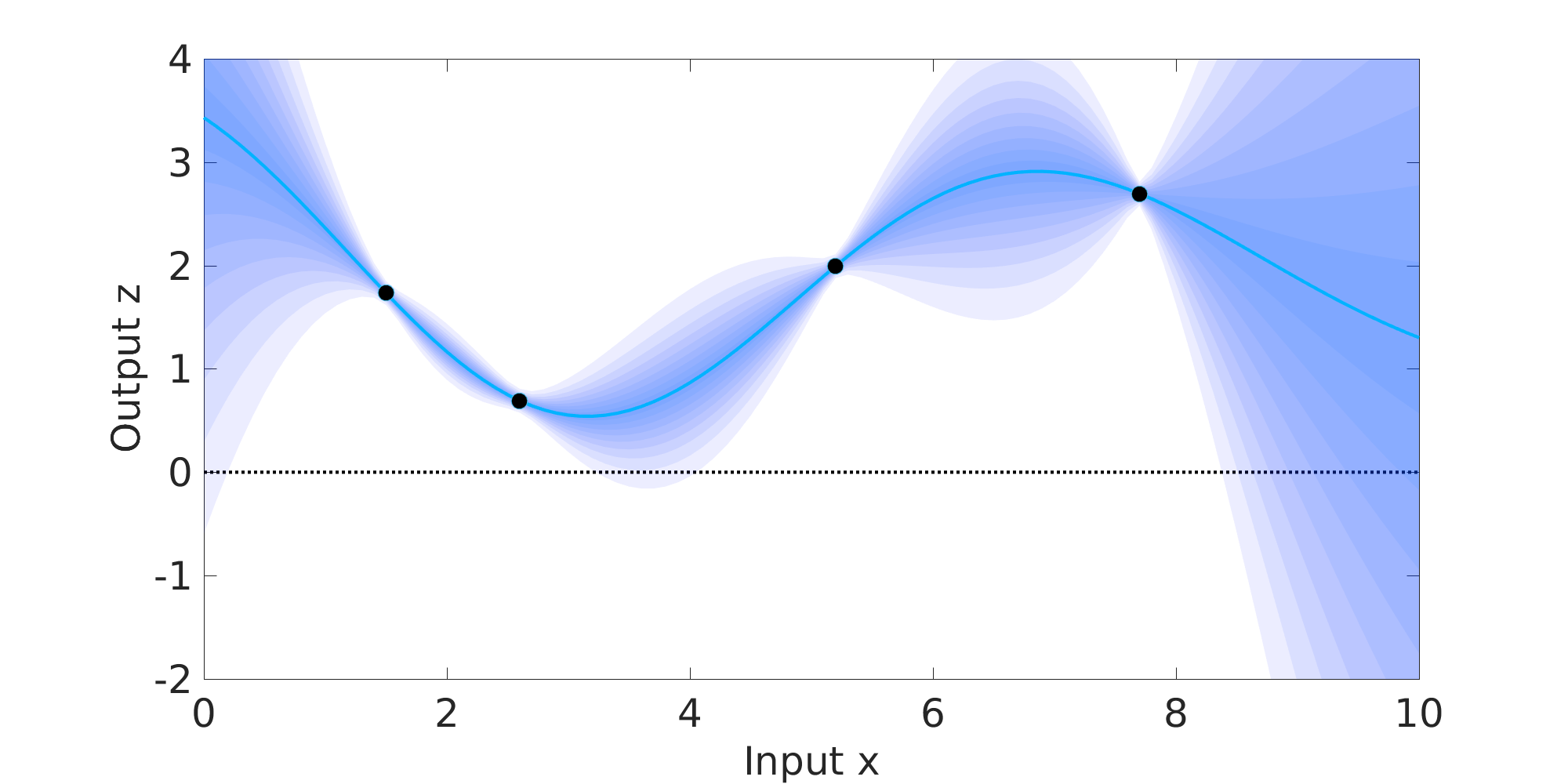}\\
      $l=0.1$}
    \end{minipage}
    \begin{minipage}{.3\textwidth}
      \centering{
      \includegraphics[width=\textwidth]{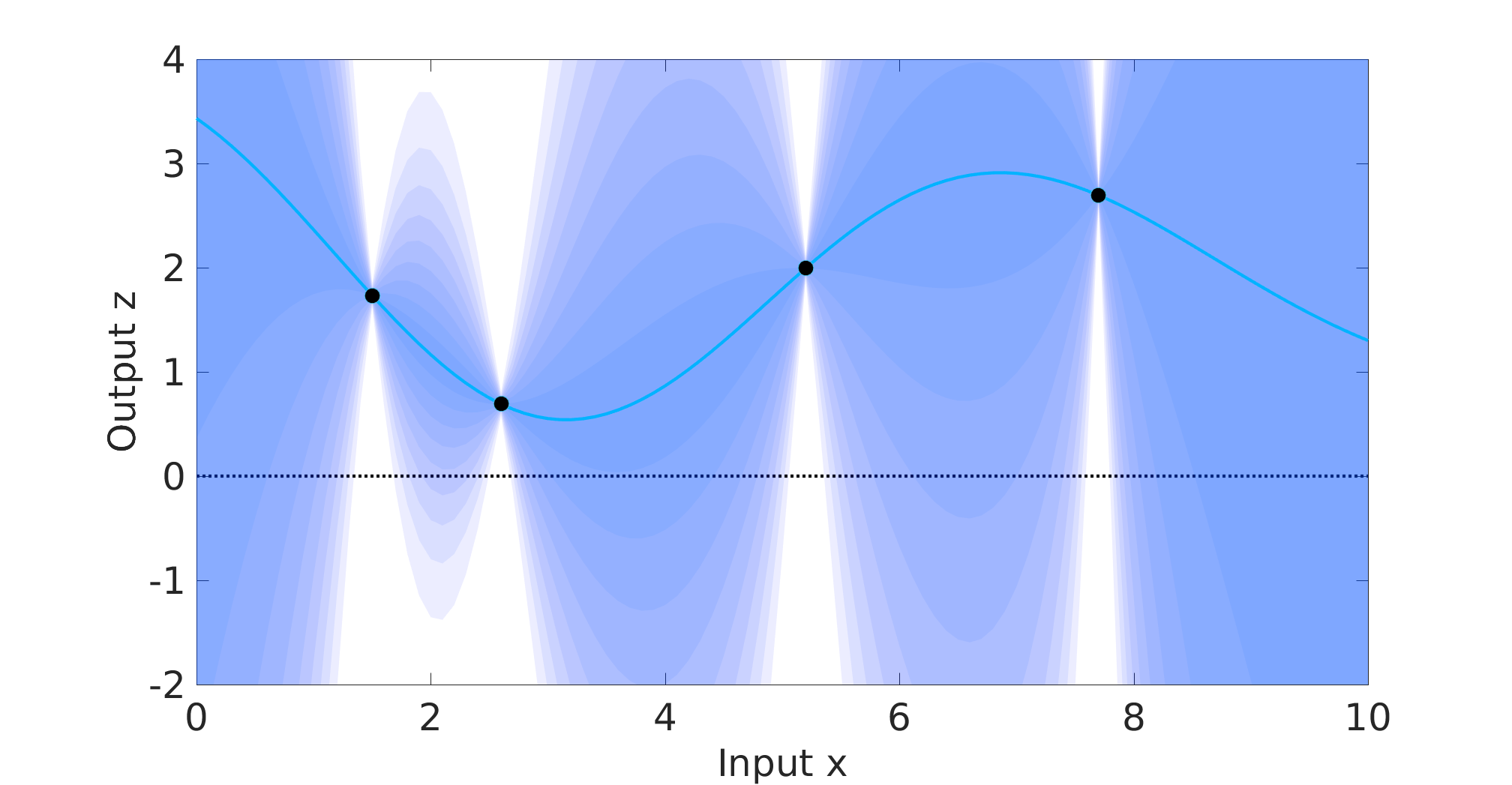}\\
      $l=0.01$}
    \end{minipage}
    \caption{ 
      Gaussian process regression
      with \it{the squared-exponential covariance kernel} 
      $k(\mathbf{x},\mathbf{x'}) = \exp(- || \mathbf{x-x'} ||^2 / 2l^2)$
      with different length scale $(l = 1, 0.1, 0.01)$.
     The black dots designate previous observations
     $D = \bigl\{\, (\mathbf{x}_{1}, f(\mathbf{x}_{1})),
     \dots, (\mathbf{x}_{t}, f(\mathbf{x}_{t})) \,\bigr\}$. 
     By the definition of Gaussian processes, 
     when one looks at the vertical section of each figure 
     at specific input $\mathbf{x}\in\mathbf{X}$, 
     the shade of blue forms a normal distribution 
     $\GP(\mu,k)(\mathbf{x})=\mathcal{N}\bigl(\mu(\mathbf{x}), k(\mathbf{x},\mathbf{x})\bigr)$.
     The center of the pipe (the light-blue line in each figure) 
     stands for the expected value $\mu(\mathbf{x})$ of the unknown function $f$; 
     the width of the pipe stands for its variance $k(\mathbf{x},\mathbf{x})$. 
    }
    \label{fig:GP}
  \end{minipage}  
\end{figure}

In this paper we follow the workflow in Algorithm~\ref{alg:falsificationByOptHighLevel}, deriving the cost function $f_{\varphi}$ in it from a Bayesian network. For the optimization step (Line~\ref{Line:algfalsificationByOptHighLevelGuess} of Algorithm~\ref{alg:falsificationByOptHighLevel}) we use \emph{Gaussian process optimization}---we follow~\cite{DBLP:journals/corr/ChenSK16a,DBLP:conf/cdc/ChenSK16,DBLP:conf/rv/Akazaki16} about this choice. It has a feature that it suggests the global shape of an unknown function; this feature turns out to be convenient for our purpose of integrating causal information in falsification. We present a brief review of the topic; see e.g.~\cite{Rasmussen:2005:GPM:1162254} for details.

\subsubsection{Gaussian Process Regression}\label{subsubsec:GPRegression}

Let $f$ be an unknown function, from a certain input domain to the set of real numbers, about which we wish to infer certain properties. (For Algorithm~\ref{alg:falsificationByOptHighLevel} we would take $f=f_{\varphi}(\mathcal{M}(\place))$). In \emph{Gaussian process regression} the shape of $f$ is estimated assuming that $f$ is a probabilistic process called a \emph{Gaussian process}. 


We start with some formal definitions. For more detail, see e.g.~\cite{Rasmussen:2005:GPM:1162254}.

\begin{mynotation}
  We let $\mathcal{N}(\boldsymbol{\mu}, \mathbf{k})$
  stand for the probability density function
  of the multivariate Gaussian distribution
  whose mean vector is $\boldsymbol{\mu}$
  and covariance matrix is $\mathbf{k}$.
\end{mynotation}

\begin{mydefinition}[Gaussian process]
  A \emph{Gaussian process} is 
  a family of probabilistic variables 
  $(\mathbf{z}_\mathbf{x})_{\mathbf{x} \in \mathbf{X}}$
  such that 
  each of its finite subset 
  $(\mathbf{z}_{\mathbf{x}_{1}}, \dots, \mathbf{z}_{\mathbf{x}_{t}})$
  has a joint Gaussian distribution.
  A Gaussian process 
  is determined by a pair $(\mu,k)$ of
  its \emph{mean function} $\mu: \mathbf{X} \to \R$ and 
  its \emph{covariance function} $k: \mathbf{X} \times \mathbf{X} \to \R$; 
  this Gaussian process is denoted by $\GP(\mu, k)$.  
  For this we have
  \begin{equation*}
    (\mathbf{z}_{\mathbf{x}_{1}}, \dots, \mathbf{z}_{\mathbf{x}_{t}})^{\top}
    \sim 
    \mathcal{N}(\boldsymbol{\mu}, \mathbf{k}) 
    \quad \text{where} \quad 
    \boldsymbol{\mu}_i = \mu(\mathbf{x}_{i})
    \quad \text{and} \quad
    \mathbf{k}_{ij} = k(\mathbf{x}_{i},\mathbf{x}_{j})
  \end{equation*}
  for each finite subset $\{\mathbf{x_1},\dotsc, \mathbf{x_t}\}$ of $\mathbf{X}$. 
  We write 
  $\GP(\mu,k)(\mathbf{x_1}, \dots, \mathbf{x_t})$ for
  the above multivariate Gaussian distribution $\mathcal{N}(\boldsymbol{\mu}, \mathbf{k})$.
\end{mydefinition}


In Fig.~\ref{fig:GP} is how an unknown function $f$ can be guessed 
by Gaussian processes.
The blue pipe designates the estimated values of the unknown function $f$:
the farther input $\mathbf{x}$ is from the observed points, 
the thicker the pipe is (that means bigger uncertainty). 

In the regression of $f$ using Gaussian processes, 
a choice of a covariance function
$k: \mathbf{X} \times \mathbf{X} \to \R$ determines \emph{smoothness} of $f$. A common template for covariance functions is the \emph{squared-exponential kernel function} 
$k_l (\mathbf{x},\mathbf{x'})\Defeq \exp(-l\cdot\|\mathbf{x}-\mathbf{x'}\|^2/2)$, where 
 $l$ is so-called the \emph{length scale} parameter.
In practice,
we pick a good length scale parameter by cross validation.
As we see in Fig.~\ref{fig:GP},
the choice of a covariance function 
yields the following tendencies in Gaussian process regression:
\begin{itemize}
 \item The bigger the distance $\|\mathbf{x}-\mathbf{x'}\|$ is, the smaller the covariance is, thus the harder it gets to estimate the value $f_\varphi(\mathbf{x})$ 
  from the observation of the value $f_\varphi(\mathbf{x'})$.
\item Covariance is smaller too when 
  the length scale parameter $l$ is bigger.
\end{itemize}

One  advantage of Gaussian process regression is that,
given a set of observations,
the \emph{posterior} process is described analytically.
Let random variables  $(\mathbf{z}_\mathbf{x})_{\mathbf{x} \in \mathbf{X}}$ obey a
prior Gaussian process  $\GP(\mu, k)$; and 
$D = \bigl\{\, (\mathbf{x}_{1}, \mathbf{z}_1),
\dotsc, (\mathbf{x}_{t}, \mathbf{z}_{t}) \,\bigr\}$ 
be a set of observations. 
Then the posterior distribution, denoted by $\GP(\mu, k; D)$, is given by 
the Gaussian process $\GP(\mu', k')$, where
\begin{align*}
  \mu'(\mathbf{x}) &= 
                     \mu(\mathbf{x}) 
                     + \mathbf{k_{\it D}(x)}
                     \mathbf{k_{\it D\it D}}^{-1}
                     \bigl(\,[\mathbf{z}_1 \dots \mathbf{z}_{t}]^\top 
                     - [\mu(\mathbf{x}_{1}) \dots \mu(\mathbf{x}_{t})]^\top\,\bigr),
  \\
  k'(\mathbf{x},\mathbf{x'}) &=
                               k'(\mathbf{x},\mathbf{x'})
                               - \mathbf{k_{\it D}(x)} \mathbf{k_{\it DD}}^{-1} \mathbf{k_{\it D}(x')}^\top.
\end{align*}
Here
$\mathbf{k_{\it D}(x)} 
= [k(\mathbf{x}_{1}, \mathbf{x}) \dots k(\mathbf{x}_{t}, \mathbf{x})]$,
and $\mathbf{k_{\it DD}}$ is a $t \times t$ matrix 
whose $i,j$-component is $k(\mathbf{x}_{i},\mathbf{x}_{j})$.
In practice,
given observed data
$D = \bigl\{\, (\mathbf{x}_{1}, f(\mathbf{x}_{1})),
\dots, (\mathbf{x}_{t}, f(\mathbf{x}_{t})) \,\bigr\}$
and a covariance kernel function $k$,
we estimate the function $f$ as $\GP(\mathbf{0}, k; D)$
where $\mathbf{0}$ denotes the function constantly zero.

\subsubsection{Gaussian Process Optimization and Acquisition Function}\label{subsec:GPPS}

Gaussian process regression allows us to predict, based on observations in $D$,  the value $f(\mathbf{x})$ for each input  $\mathbf{x}$ as a normal distribution $\GP(\mu,k)(\mathbf{x})$. To complete an optimization scenario, 
we wish to pick 
a candidate $\mathbf{x} \in \mathbf{X}$ for which $f(\mathbf{x})$ is small.


It is well-known that, for such choice, a balance is important
between \emph{exploration} (i.e.\ bias toward a bigger variance $k(\mathbf{x},\mathbf{x})$) and \emph{exploitation} (bias toward a smaller expected value $\mu(\mathbf{x})$). A criterion for this balance is called an \emph{acquisition function}---we pick $\mathbf{x}$ at which the acquisition function is minimum. Assuming that 
an acquisition function
$\psi(\mathbf{x}; \GP(\mu,k))$ has been fixed, 
the whole procedure for  Gaussian process optimization 
can be described as in Algorithm~\ref{algo:GPOptimization}.
Note that, in Line~\ref{line:GPOptChooseNextInput} of Algorithm~\ref{algo:GPOptimization}, we usually employ another optimization solving method (such as simulated annealing).



\auxproof{\color{red}
\begin{myremark}
  Note that,
  in falsification for CPSs,
  computing the acquisition function is
  relatively easier than
  obtaining the output of the system $S(\mathbf{x})$
  because the system $S$ is complex.
  As we see in \S{}\ref{sec:Experiment},
  even if our example systems are much simpler than
  the models in practical use,
  our experimental result shows that
  this ``replacement of the optimization problem''
  is still effective.
\end{myremark}
\color{black}
}

In falsification,
our goal would be to find $\mathbf{x}$ such that
$f(\mathbf{x}) < 0$. 
As a natural choice of acquisition functions,
we focus on the following probability in this paper.
\begin{mydefinition}[Probability of Satisfaction]\label{def:PSat}
  \begin{equation}
    \label{eq:PSat}
    \psi(\mathbf{x}; \GP(\mu,k)) \;\Defeq \;
    \Prob_{\GP(\mu,k)}(f(\mathbf{x}) > 0)
  \end{equation}
  Here
  $\Prob_{\GP(\mu,k)}(f(\mathbf{x}) > c)$
  is an abbreviation of 
  $\Prob(f(\mathbf{x}) > c 
  \mid f(\mathbf{x}) \sim \GP(\mu,k)(\mathbf{x}))$.
\end{mydefinition}

\begin{figure}[tbp]
  \centering
  \begin{minipage}{\textwidth}
    \centering
    \begin{minipage}{.22\textwidth}
      \centering
      \includegraphics[width=\textwidth]{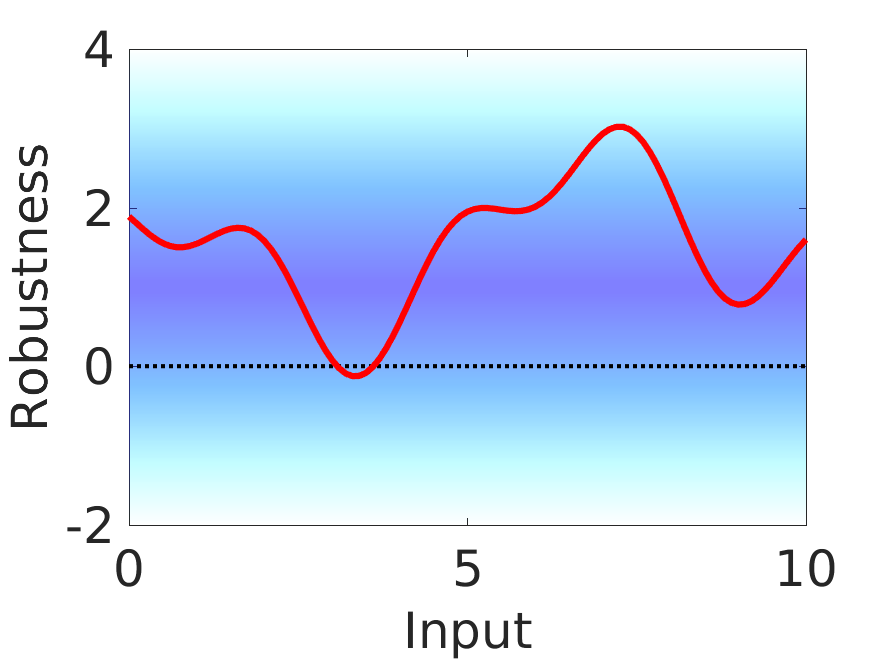}\\
      time $t=1$\\
      (Next: $\mathbf{x}_1=5.5$)\\
    \end{minipage}
    \begin{minipage}{.22\textwidth}
      \centering
      \includegraphics[width=\textwidth]{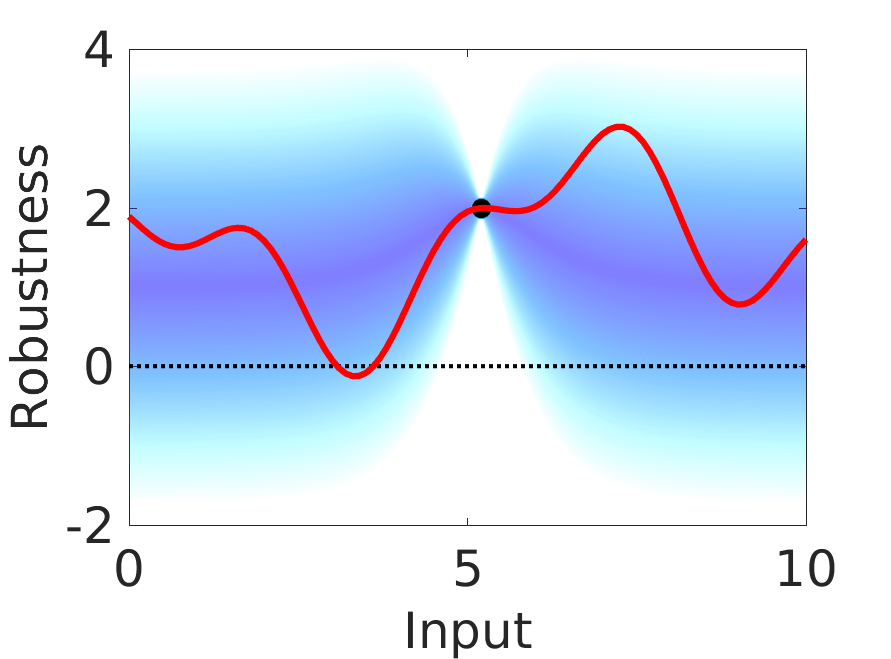}\\
      $t=2$\\
      (Next: $\mathbf{x}_2=0$)\\
    \end{minipage}
    \begin{minipage}{.22\textwidth}
      \centering
      \includegraphics[width=\textwidth]{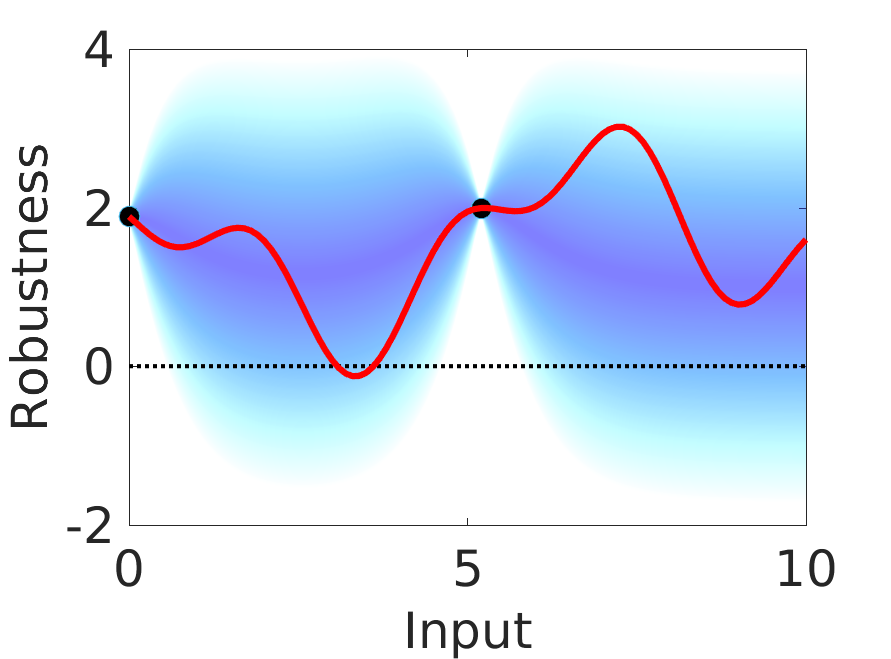}\\
      $t=3$\\
      (Next: $\mathbf{x}_3=10$)\\
    \end{minipage}
    \begin{minipage}{.22\textwidth}
      \centering
      \includegraphics[width=\textwidth]{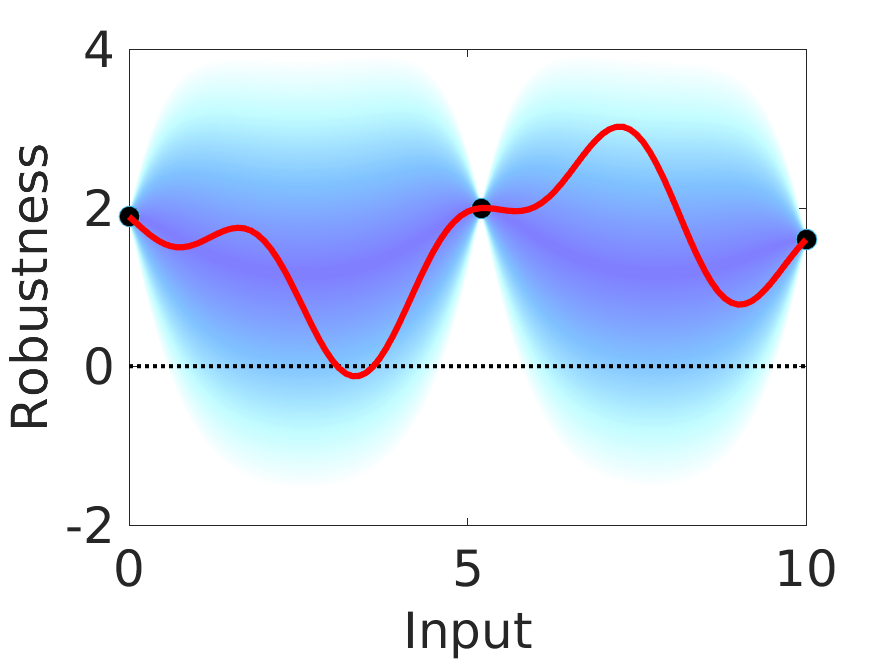}\\
      $t=4$\\
      (Next: $\mathbf{x}_4=2.6$)\\
    \end{minipage}
    \begin{minipage}{.22\textwidth}
      \centering
      \includegraphics[width=\textwidth]{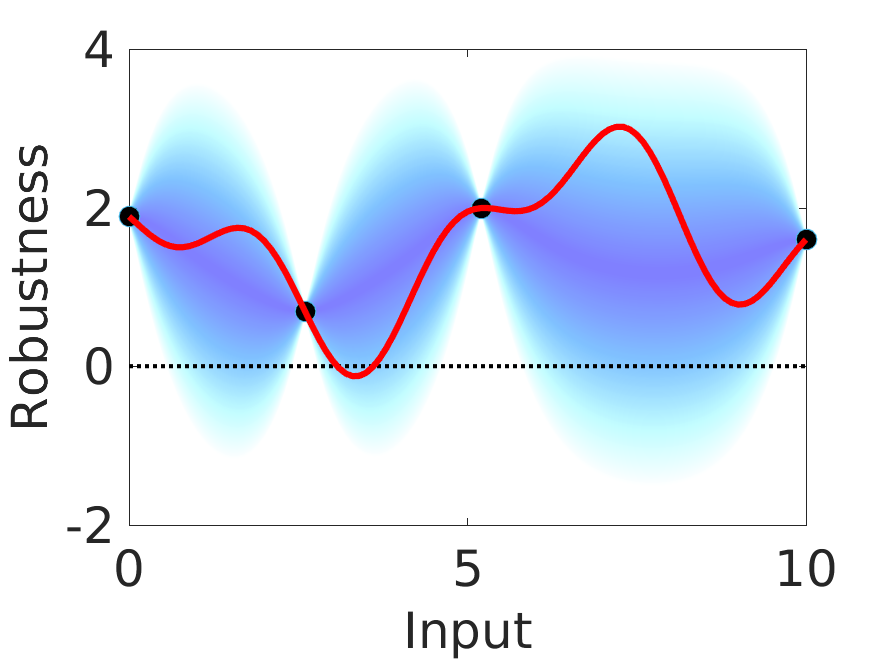}\\
      $t=5$\\
      (Next: $\mathbf{x}_5=7.8$)\\
    \end{minipage}
    \begin{minipage}{.22\textwidth}
      \centering
      \includegraphics[width=\textwidth]{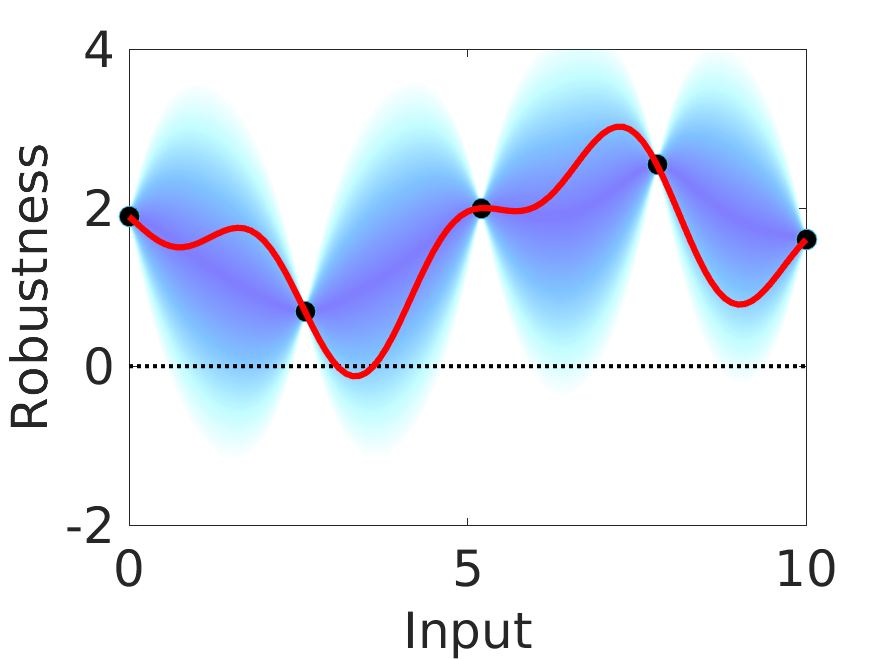}\\
      $t=6$\\
      (Next: $\mathbf{x}_6=1.8$)\\
    \end{minipage}
    \begin{minipage}{.22\textwidth}
      \centering
      \includegraphics[width=\textwidth]{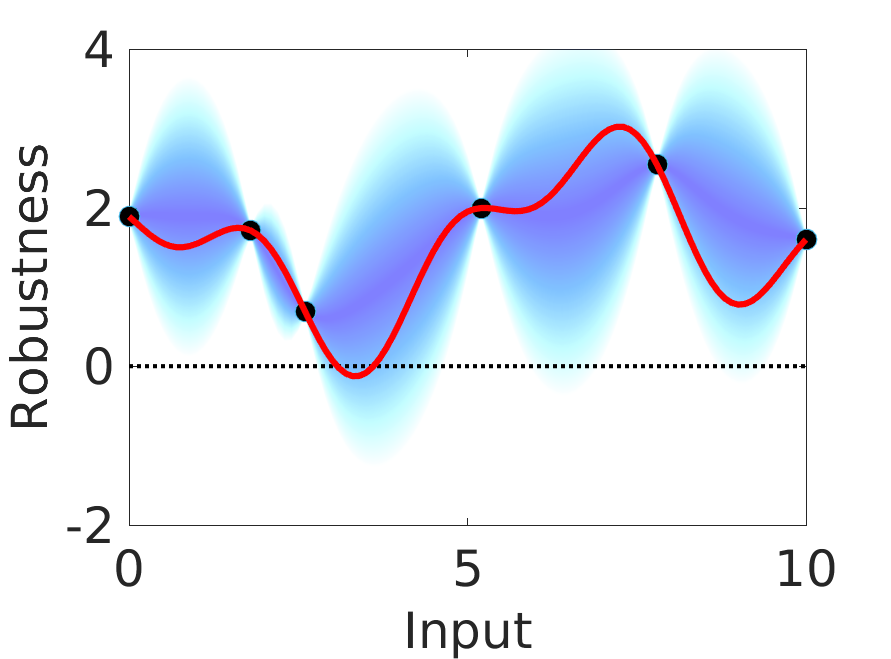}\\
      $t=7$\\
      (Next: $\mathbf{x}_7=3.3$)\\
    \end{minipage}
    \begin{minipage}{.22\textwidth}
      \centering
      \includegraphics[width=\textwidth]{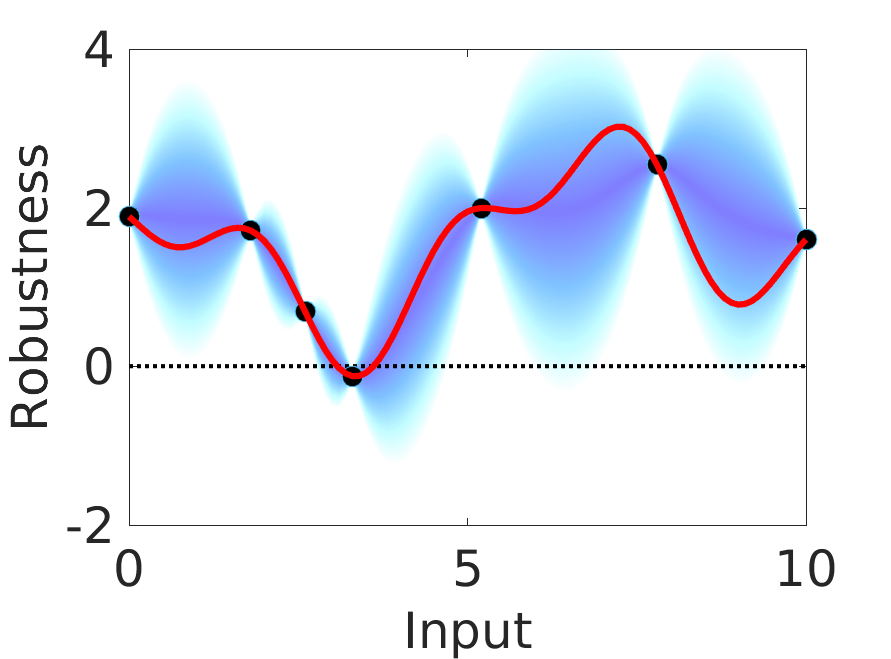}\\
      $t=8$\\
      (falsified)
    \end{minipage}
    \caption{
       Illustration of the $\GPPS$ algorithm.
      In each figure,
      the red line
      is for the unknown function $f$ to  minimize; and
      the blue cloud is
      the  Gaussian process estimation $\GP(\mu,k)$ of $f$.
      At time $t=1$ the input  is chosen randomly (say  $\mathbf{x}_{1} = 5.5$).
      Subsequently
      we pick the point $\mathbf{x}_{t}$ 
      that minimizes the probability
      $\Prob_{\GP(\mu,k)}(f(\mathbf{x}) > 0)$.
      We observe that, as the algorithm proceeds, the estimate of $f$ becomes finer too.
    }
    \label{fig:GPOptimization}
  \end{minipage}  
\end{figure}

We write $\GPPS$ for Algorithm~\ref{algo:GPOptimization} under
$\psi$ as an acquisition function;
Fig.~\ref{fig:GPOptimization} illustrates how it works.

The acquisition functions we will use are extension of this $\psi$. 
We note, however, that this acquisition function
is commonly known as ``pure and impractical'' 
in the field of Gaussian process optimization.
More sophisticated acquisition functions that are known include
\emph{probability improvement},
\emph{expected improvement} \cite{mockus1989bayesian},
\emph{upper confidence bound} \cite{DBLP:conf/icml/SrinivasKKS10} 
and so on. 
At the time of writing it is not clear 
how these acquisition functions can be used 
as part of our framework in~\S{}\ref{sec:Algorithm}. 

\begin{algorithm}[tbp]
  \caption{Gaussian process optimization $\mathsf{GPOptimization}(D,k,\psi)$}
  \label{algo:GPOptimization}
  \begin{algorithmic}[1]
    \INPUT 
    a covariance function
    $k: \mathbf{X} \times \mathbf{X} \to \R$;
    an acquisition function $\psi$; and
    an initial data set $D = \{(\mathbf{x}'_1, f(\mathbf{x}'_1)), \dots, (\mathbf{x}'_s, f(\mathbf{x}'_s))\}$
     \OUTPUT input $\mathbf{x} \in \mathbf{X}$ for which $f(\mathbf{x})$ is small
    \For{ $t = 1, 2,\dots$ }
    \State $\GP(\mu', k') 
    = \GP(\mathbf{0}, k; D)$;
    \Comment Estimate the unknown function $f$
    \State \label{line:GPOptChooseNextInput}
    $\mathbf{x}_{t} = \Argmin_{\mathbf{x \in X}} 
    \psi(\mathbf{x}; \GP(\mu', k'))$;
    \Comment Choose new sample input
    \State $D = D \cup \{(\mathbf{x}_{t}, f(\mathbf{x}_{t}))\}$;
    \Comment Observe the corresponding output
    \EndFor
   \State\Return $\mathbf{x}_{t}$
  \end{algorithmic}
\end{algorithm}

\section{Causality in Falsification: Further Examples}\label{sec:BayesianNet}
In addition to Example~\ref{ex:counter},  we shall exhibit two more examples of falsification problems; for each, we introduce a Bayesian network that encodes suitable causal information, too. The latter will be exploited in our causality-aided algorithm in~\S{}\ref{sec:Algorithm}.

\begin{figure}[tbp]
  \begin{minipage}{.4\textwidth}
    \centering
    \begin{minipage}{.7\textwidth}
      \centering
      \begin{algorithmic}
        \INPUT $i_1, \dots i_4 \in [0,1]$
        \OUTPUT $x_1(t),\dotsc,x_{4}(t)$ 
        
        for each $t\in\Rnn$
        \State $x_1(t) = \sin(1.1t + i_1)$;
        \State $x_2(t) = \sin(1.2t + i_2)$;
        \State $x_3(t) = \sin(1.3t + i_3)$;
        \State $x_4(t) = \sin(1.4t + i_4)$;
      \end{algorithmic}
    \end{minipage}
    \caption{System model for \S{}\ref{subsec:sineSystem}}
    \label{fig:sineSystem}
  \end{minipage}
  \qquad
   \begin{minipage}{.5\textwidth}
  \begin{minipage}{\textwidth}
    \scalebox{.9}{
     \begin{minipage}{.54\textwidth}\label{fig:Example2}
      \begin{tikzpicture}[
        node distance=1cm and 0cm,
        mynode/.style={draw,ellipse,text width=1.7cm,align=center}
        ]
        \node[mynode] (r) {$\varphi$: \\$\BoxOp{}(p_{12}\vee p_{34})$};
        \node[mynode, text width=1.5cm,
        above left=1.5cm and -1.5cm of r] (c1)
        {$\varphi_{12}$: $\BoxOp{}p_{12}$};
        \node[mynode, text width=1.5cm, 
        above right=1.5cm and 1cm of r] (c2)
        {$\varphi_{34}$: $\BoxOp{}p_{34}$};
        \node[rectangle callout,draw,inner sep=2pt,
        callout absolute pointer=(c1.north),
        above= 0.5cm of c1](b1){
          \begin{minipage}{7.5em}
            \centering
            \begin{tabular}{ p{0.5cm}||cc }
              \qquad
              & $ \ttrue$
              & $ \ffalse$
              \\ \hhline{=||==}
              & 0.9 & 0.1 \\
            \end{tabular}
            \end{minipage}};
        \node[rectangle callout,draw,inner sep=2pt,
        callout absolute pointer=(c2.north),
        above= 0.5cm of c2](b2){
          \begin{minipage}{7.5em}
            \centering
            \begin{tabular}{ p{0.5cm}||cc }
              \qquad
              & $ \ttrue$
              & $ \ffalse$
              \\ \hhline{=||==}
              & 0.9 & 0.1 \\
            \end{tabular}
            \end{minipage}};
        \node[rectangle callout,draw,inner sep=2pt,
        callout absolute pointer=(r.east),
        right= 0.5cm of r](b){
          \begin{minipage}{10.0em}
            \centering
            \begin{tabular}{ cc||cc }
              $\varphi_{12}$
              & $\varphi_{34}$
              & $ \ttrue$
              & $ \ffalse$ \\ \hhline{==||==}
              $\ttrue$ & $\ttrue$ & 1 & 0 \\
              $\ttrue$ & $\ffalse$ & 1 & 0 \\
              $\ffalse$ & $\ttrue$ & 1 & 0 \\
              $\ffalse$ & $\ffalse$ & 0.9 & 0.1 \\
            \end{tabular}
            \end{minipage}};
        \path
        (c1) edge[-latex] (r)
        (c2) edge[-latex] (r);
      \end{tikzpicture}
      where $\bigg(
      \begin{array}{c}
        p_{12} \equiv x_1 < 0.99 \vee x_2 < 0.99\\
        p_{34} \equiv x_3 < 0.99 \vee x_4 < 0.99        
      \end{array}\bigg)$    
   \end{minipage}
   }
   \end{minipage}
        \caption{Bayesian network for \S{}\ref{subsec:sineSystem}}
        \label{fig:sineBayes}
\end{minipage}

\vspace{1.5em}
\begin{minipage}{\textwidth}
     \centering
       \scalebox{.85}{
    \begin{minipage}{\linewidth}
      \begin{tabular}{ccc||p{0.8cm}}
        \multicolumn{4}{c}{$\Prob_{\mathcal{B}}(-)$}\\
        $\varphi_1$ 
     & $\varphi_2$ 
     & $\varphi$ & \\ \hhline{===||=}
        $\ttrue$ & $\ttrue$ & $\ttrue$ & 0.81 \\ 
        $\ttrue$ & $\ttrue$ & $\ffalse$ & 0 \\ 
        $\ttrue$ & $\ffalse$ & $\ttrue$ & 0.09 \\ 
        $\ttrue$ & $\ffalse$ & $\ffalse$ & 0 \\ 
        $\ffalse$ & $\ttrue$ & $\ttrue$ & 0.09 \\ 
        $\ffalse$ & $\ttrue$ & $\ffalse$ & 0 \\ 
        $\ffalse$ & $\ffalse$ & $\ttrue$ & 0.009 \\ 
        $\ffalse$ & $\ffalse$ & $\ffalse$ & 0.001 \\ 
      \end{tabular}\quad
      \begin{tabular}{ccc||p{0.8cm}}
        \multicolumn{4}{c}{$\Prob_{\mathcal{B}}(- \mid \sem{\varphi} = \ffalse)$}\\
        $\varphi_1$ 
     & $\varphi_2$ 
     & $\varphi$ & \\ \hhline{===||=}
        $\ttrue$ & $\ttrue$ & $\ttrue$ & 0 \\ 
        $\ttrue$ & $\ttrue$ & $\ffalse$ & 0 \\ 
        $\ttrue$ & $\ffalse$ & $\ttrue$ & 0 \\ 
        $\ttrue$ & $\ffalse$ & $\ffalse$ & 0 \\ 
        $\ffalse$ & $\ttrue$ & $\ttrue$ & 0 \\ 
        $\ffalse$ & $\ttrue$ & $\ffalse$ & 0 \\ 
        $\ffalse$ & $\ffalse$ & $\ttrue$ & 0 \\ 
        $\ffalse$ & $\ffalse$ & $\ffalse$ & 1 \\ 
      \end{tabular}
      \quad
      \begin{tabular}{c||p{1.2cm}}
        \multicolumn{2}{c}{$\Prob_{\mathcal{B}}(-)$}\\ \hhline{=||=}
        $\varphi_{12} = \ttrue$ & 0.9 \\ 
        $\varphi_{12} = \ffalse$ & 0.1 \\  \hline
        $\varphi_{34} = \ttrue$ & 0.9 \\ 
        $\varphi_{34} = \ffalse$ & 0.1 \\ \hline
        $\varphi = \ttrue$ & 0.999 \\ 
        $\varphi = \ffalse$ & 0.001 \\ 
      \end{tabular}\quad
      \begin{tabular}{c||p{0.8cm}}
        \multicolumn{2}{c}{$\Prob_{\mathcal{B}}(-\mid \sem{\varphi}=\ffalse)$}\\
        \hhline{=||=}
        $\varphi_{12} = \ttrue$ & 0 \\ 
        $\varphi_{12} = \ffalse$ & 1 \\ \hline
        $\varphi_{34} = \ttrue$ & 0 \\ 
        $\varphi_{34} = \ffalse$ & 1 \\ \hline
        $\varphi = \ttrue$ & 0 \\ 
        $\varphi = \ffalse$ & 1 \\ 
      \end{tabular}
    \end{minipage}
    }
      \caption{ Unconditional/conditional joint distributions 
        in the Bayesian network of Fig.~\ref{fig:sineBayes}}
      \label{fig:sineJointProb}
\end{minipage}

\vspace{1.5em}
\begin{minipage}{.45\textwidth}
\scalebox{.9}{     \begin{minipage}{.54\linewidth}\label{fig:Example3}
      \begin{tikzpicture}[
        node distance=1cm and 0cm,
        mynode/.style={draw,ellipse,text width=1.7cm,align=center}
        ]
        \node[mynode] (r) {$\varphi$
 \\$\cong\varphi_v \wedge \varphi_\omega$
};
        \node[mynode, text width=1.8cm,
        above left=1.5cm and -1.5cm of r] (c1)
        {$\varphi_v$: $\BoxOp{} v < 120$};
        \node[mynode, text width=2.0cm, 
        above right=1.5cm and 1cm of r] (c2)
        {$\varphi_{\omega}$: $\BoxOp{}\omega < 4780$};
        \node[rectangle callout,draw,inner sep=2pt,
        callout absolute pointer=(c1.north),
        above= 0.5cm of c1](b1){
          \begin{minipage}{7.5em}
            \centering
            \begin{tabular}{ p{0.1cm}||cc }
              \qquad
              & $ \ttrue$
              & $ \ffalse$
              \\ \hhline{=||==}
              & 0.99 & 0.01 \\
            \end{tabular}
            \end{minipage}};
        \node[rectangle callout,draw,inner sep=2pt,
        callout absolute pointer=(c2.north),
        above= 0.5cm of c2](b2){
          \begin{minipage}{7.0em}
            \centering
            \begin{tabular}{ p{0.1cm}||cc }
              \qquad
              & $ \ttrue$
              & $ \ffalse$
              \\ \hhline{=||==}
              & 0.9 & 0.1 \\
            \end{tabular}
            \end{minipage}};
        \node[rectangle callout,draw,inner sep=2pt,
        callout absolute pointer=(r.east),
        right= 0.5cm of r](b){
          \begin{minipage}{7.5em}
            \centering
            \begin{tabular}{ cc||cc }
              $\varphi_{v}$
              & $\varphi_{\omega}$
              & $ \ttrue$
              & $ \ffalse$ \\ \hhline{==||==}
              $\ttrue$ & $\ttrue$ & 1 & 0 \\
              $\ttrue$ & $\ffalse$ & 0 & 1 \\
              $\ffalse$ & $\ttrue$ & 0 & 1 \\
              $\ffalse$ & $\ffalse$ & 0 & 1 \\
            \end{tabular}
            \end{minipage}};
        \path
        (c1) edge[-latex] (r)
        (c2) edge[-latex] (r);
      \end{tikzpicture}
    \end{minipage}}
   \caption{Bayesian network for \S{}\ref{subsec:at}}
   \label{fig:atBayes}
\end{minipage}
\quad
\begin{minipage}{.5\textwidth}
  \scalebox{.9}{
      \begin{minipage}{\textwidth}
      \begin{tabular}{ccc||p{0.8cm}}
        \multicolumn{4}{c}{$\Prob_{\mathcal{B}}(-)$}\\
        $\varphi_v$ 
        & $\varphi_w$ 
        & $\varphi$ & \\ \hhline{===||=}
        $\ttrue$ & $\ttrue$ & $\ttrue$ & 0.891 \\ 
        $\ttrue$ & $\ttrue$ & $\ffalse$ & 0 \\ 
        $\ttrue$ & $\ffalse$ & $\ttrue$ & 0 \\ 
        $\ttrue$ & $\ffalse$ & $\ffalse$ & 0.099 \\ 
        $\ffalse$ & $\ttrue$ & $\ttrue$ & 0 \\ 
        $\ffalse$ & $\ttrue$ & $\ffalse$ & 0.009 \\ 
        $\ffalse$ & $\ffalse$ & $\ttrue$ & 0 \\ 
        $\ffalse$ & $\ffalse$ & $\ffalse$ & 0.001 \\ 
      \end{tabular}\quad
      \begin{tabular}{ccc||p{0.8cm}}
        \multicolumn{4}{c}{$\Prob_{\mathcal{B}}(-\mid\sem{\varphi}=\ffalse)$}\\
        $\varphi_v$ 
        & $\varphi_w$ 
        & $\varphi$ & \\ \hhline{===||=}
        $\ttrue$ & $\ttrue$ & $\ttrue$ & 0 \\ 
        $\ttrue$ & $\ttrue$ & $\ffalse$ & 0 \\ 
        $\ttrue$ & $\ffalse$ & $\ttrue$ & 0 \\ 
        $\ttrue$ & $\ffalse$ & $\ffalse$ & 0.908 \\ 
        $\ffalse$ & $\ttrue$ & $\ttrue$ & 0 \\ 
        $\ffalse$ & $\ttrue$ & $\ffalse$ & 0.083 \\ 
        $\ffalse$ & $\ffalse$ & $\ttrue$ & 0 \\ 
        $\ffalse$ & $\ffalse$ & $\ffalse$ & 0.009 \\ 
      \end{tabular}\\
      \begin{tabular}{c||p{1.2cm}}
        \multicolumn{2}{c}{$\Prob_{\mathcal{B}}(-)$}\\ \hhline{=||=}
        $\varphi_v = \ttrue$ & 0.99 \\ 
        $\varphi_v = \ffalse$ & 0.01 \\  \hline
        $\varphi_w = \ttrue$ & 0.9 \\ 
        $\varphi_w = \ffalse$ & 0.1 \\ \hline
        $\varphi = \ttrue$ & 0.891 \\ 
        $\varphi = \ffalse$ & 0.109 \\ 
      \end{tabular}\quad
      \begin{tabular}{c||p{0.8cm}}
        \multicolumn{2}{c}{$\Prob_{\mathcal{B}}(-\mid\sem{\varphi}=\ffalse)$}\\
        \hhline{=||=}
        $\varphi_v = \ttrue$ & 0.908 \\ 
        $\varphi_v = \ffalse$ & 0.092 \\ \hline
        $\varphi_w = \ttrue$ & 0.083 \\ 
        $\varphi_w = \ffalse$ & 0.917 \\ \hline
        $\varphi = \ttrue$ & 0 \\ 
        $\varphi = \ffalse$ & 1 \\ 
      \end{tabular}
    \end{minipage}
   }
   \caption{Unconditional/conditional joint distributions 
        in the Bayesian network of Fig.~\ref{fig:atBayes}}
   \label{fig:atJointProb}
\end{minipage}
\end{figure}

\subsection{Example Model 2: Coincidental Sine Waves}\label{subsec:sineSystem}
Let us consider the model in Fig.~\ref{fig:sineSystem}.
In this simple model
there are four sine waves $x_1(t), \dots, x_4(t)$ of different frequency,
and we pick their initial phases $i_1, \dots, i_4$
as input of the system.

As a specification,
we pick the following formula---it is falsified
when the peaks of four sine waves correspond.

\begin{equation}
  \label{eq:Example2Spec}
  \varphi \;\equiv\; \BoxOp{[0,10]}(\bigvee_{i=1,\dotsc,4} x_i < 0.99 )
\end{equation}

We see that
falsifying $\varphi$ with pure random sampling is difficult
because
$\varphi$ is false 
only in rare cases.
What is worth, the (conventional) robustness of $\varphi$ 
does not always guide us to the counterexamples.
\begin{myexample}
Let us consider 
the subformula $\bigvee_{i=1,\dots,4} x_i < 0.99$.
When we compare the values $(x_1, x_2, x_3, x_4) = (1,1,1,0)$ and $(0,0,0,1)$,
we could say the former is ``closer'' to falsifying the subformula---the 
three out of four sine waves simultaneously at a peak.  
However, these robustness values are the same $0.99$ in both cases.
\end{myexample}


In this case,
we sometimes divide the difficulty into small pieces---first 
get $x_{1}$ and $x_{2}$ simultaneously at a peak; then get $x_{3}$ and $x_{4}$ at a peak; finally, try to make them synchronize.
Let us introduce formulas $\varphi_{12}$ and $\varphi_{34}$
such that falsifying them means 
matching the peak of $x_1, x_2,$ and $x_3, x_4$ respectively.
Decomposing $\varphi$ into $\varphi_{12}$ and $\varphi_{34}$
might help us in falsification 
for the following reasons.
\begin{itemize}
\item The small formulas $\varphi_{12}$ and $\varphi_{34}$ 
  are much easier to falsify compared to $\varphi$. 
\item Moreover, the robustness mapping 
  $f_{\varphi_{12}}(\mathcal{M}(\place))$ and $f_{\varphi_{34}}(\mathcal{M}(\place))$
  have much simpler dynamics than the one of the original specification $\varphi$,
  so the Gaussian process regression for the small formulas 
  tend to work better
  than the one for $\varphi$.
\end{itemize}

The Bayesian network $\mathcal{B}$ in Fig.~\ref{fig:sineBayes} is devised 
to express this intuition.
For example,
the formula $\varphi$ is true with probability $1$
when either $\varphi_{12}$ or $\varphi_{34}$,
otherwise $\varphi$ becomes false with small probability $0.1$.
As shown in Fig.~\ref{fig:sineJointProb},
the conditional joint distribution
$\Prob_{\mathcal{B}}(-\mid \sem{\varphi}=\ffalse)$
tells us the fact that $\varphi$ is false
only if both $\varphi_{12}$ and $\varphi_{34}$ are false.

\subsection{Example Model 3: Automatic Transmission}\label{subsec:at}
The last example is the automatic transmission model 
from the benchmark of temporal logic verification~\cite{HoxhaAF14arch1}.
This model is 
still miniature,
but an elaborate mimicry of the systems in the real world
hence suitable for our purpose.
 
As a specification $\varphi$ to falsify,
we use the following formula. It is taken from~\cite{HoxhaAF14arch1} (it is $\varphi_{2}^{AT}$ there).
\begin{align*}
 \BoxOp{} (v < 120 \wedge \omega < 4780)
\end{align*}
Here the variable $v$ and $\omega$
stand for
the speed of the vehicle
and the angular velocity of the engine rotation respectively.

We know that
we can falsify $\varphi$
either by violating the speed limit ($v < 120$) or
the engine rotation limit ($\omega < 4780$).
In this model,
$\omega$ takes the values in the range around $[0, 4800]$
while $v$ does around $[0, 120]$. Note that their scales are very different: 
hence in the most of the cases,
the robustness of the $\omega$-component is likely to be
shadowed by the one of the $v$-component.
As a consequence,
we  expect that
conventional falsification solver
only try to falsify
by the violation of the speed limit $v < 120$. 

The Bayesian network annotation
is also effective
in such a situation.
That is,
we can add the information about
``which is more likely to happen, 
the violation of the speed and the rotation limit.''
(In actual deployment such insights will be provided by engineers' domain knowledge.)
Let assume that
the probabilities of 
the violation of the speed and the rotation limit
are $0.01$ and $0.1$ respectively.
This information is expressed in Fig.~\ref{fig:atBayes}, 
where the conditional probabilities for $\varphi$ 
simply encode logical relationship 
(note that $\varphi$ is semantically equivalent to $\varphi_v \wedge \varphi_\omega$) 
however, the probabilities at leaves reflect the above insight. 

\begin{myremark}
  In \S{}\ref{subsec:sineSystem} and \S{}\ref{subsec:at},
  as an indicator of robustness,
  we employed the (space) robust semantics of $\STL$ 
  in \cite{DBLP:conf/formats/DonzeM10}
  and shown that
  it is not sensitive enough for some falsification scenarios.
  In contrast to \cite{DBLP:conf/formats/DonzeM10},
  the metric-based robustness of $\MITL$ in \cite{DBLP:journals/tcs/FainekosP09}
  has a degree of freedom
  to capture the lacked notions.
  For example in \S{}\ref{subsec:at},
  we could solve the falsification problem more efficiently
  if we could re-scale $v$ and $\omega$ appropriately,
  and this re-scaling is nothing but the defining the metric space in \cite{DBLP:journals/tcs/FainekosP09}. 
  However,
  defining such a metric space itself is challenging 
  and needs expert's domain knowledge---similarly as our framework needs suitable causal information. 
  We expect that
  our causality-aided framework is a viable option
  compare to finding a suitable metric.
\end{myremark}



\section{Falsification with Causality Annotation}\label{sec:Algorithm}
Given the backgrounds in~\S{}\ref{sec:bck} and the examples in~\S{}\ref{sec:BayesianNet}, we are now ready to ask the question:
\emph{given a falsification problem
and a Bayesian network annotation about causality,
what cost function should we optimize?}
In this section,
we will give some answers
to the question
by lifting up 
the conventional notion of acquisition functions
which we reviewed in \S{}\ref{subsec:GPOptimization}
to the multi-formula setting.

Consider one of the Bayesian networks that we have seen in the paper. 
Let $\mathcal{B}$ denote  the Bayesian network; and
let $\Phi = \{\varphi_1, \dots, \varphi_N\}$ be 
the set of formulas that appear there.
Now assume that we are running the Gaussian regression
not only for $f_{\varphi}=\sem{\mathcal{M}(\place), \varphi}$ 
but also $f_{\varphi_{i}}=\sem{\mathcal{M}(\place), \varphi_{i}}$ 
for all the formulas $\varphi_{i}$ in the Bayesian network. 

The regression result for $f_{\varphi_{i}}$
gives us 
the probabilistic ``forecast''
of the truth values assignment 
of the formulas $\Theta \in 2^{\Phi}$ as follows.
\begin{mynotation}
  Let $\GP(\mu_i,k_i)\sim f_{\varphi_{i}}$ be our estimate for $ f_{\varphi_{i}}$; 
  we can use this data to estimate the probability of obtaining $\Theta$ 
  as the truth assignment, under an input value $\mathbf{x}$. Precisely:
  let $\Theta$ be the assignment
  $(\varphi_1 = \theta_1, \dots, \varphi_N = \theta_N)$
  where $\theta_i \in \{\ttrue, \ffalse\}$;
  then
  \begin{equation}
    \label{eq:ProbabilityGP}
    \Prob_{\GP(\mathbf{x})}(\Theta)
    \;\Defeq\;
    \Prob_{\GP(\mu_1,k_1)}\bigl(f_{\varphi_1}(\mathbf{x}) \;R_1\; 0\bigr)
    \cdots
    \Prob_{\GP(\mu_N,k_N)}\bigl(f_{\varphi_N}(\mathbf{x}) \;R_N\; 0\bigr),
  \end{equation}
  where $R_i$ is $>$ if $\theta_1 = \ttrue$, and  $<$ otherwise.
\end{mynotation}

\subsection{KL Divergence based acquisition function}\label{subsec:Algorithm}
Recall the scenario in \S{}\ref{subsec:sineSystem}---from the 
conditional joint distribution
$\Prob_{\mathcal{B}}(-\mid \sem{\varphi}=\ffalse)$,
we see that 
the both small formulas $\varphi_{12}$ and $\varphi_{34}$ 
also should be false
to synchronize all the peaks of the sine waves.

Inspired from the above example,
we propose the following criteria
to choose the next candidate $\mathbf{x}$ as falsifying input.

\begin{mydefinition}
  [An acquitision function $\psi_{\mathcal{B}}(\mathbf{x})$]
  \label{def:KLDist}
  \begin{equation*}
    \mathbf{x}
    = \Argmin_{\mathbf{x}} \psi_{\mathcal{B}}(\mathbf{x})
    \; \text{where} \;
    \psi_{\mathcal{B}}(\mathbf{x}) = 
    \KLDist\bigg(
    \Prob_{\mathcal{B}}(\Theta \mid \sem{\varphi} = \ffalse) 
    \,\bigg|\bigg|\,
    \Prob_{\GP(\mathbf{x})}(\Theta)
    \bigg)
  \end{equation*}
\end{mydefinition}
Here $\KLDist$ is the \emph{Kullback Leibler divergence}---a measure
of the difference between two probabilistic distributions.
Intuitively,
with this criteria,
we pick the next input $\mathbf{x}$
with which the probabilistic forecast $\Prob_{\GP(\mathbf{x})}(\Theta)$
by regression becomes ``closer to the conditional joint distribution
$ \Prob_{\mathcal{B}}(\Theta \mid \sem{\varphi} = \ffalse)$''.

\begin{myexample}
  Let us consider the sine waves model in \S{}\ref{subsec:sineSystem}.
  From simple calculation,
  we see that the acquisition function $\psi_{\mathcal{B}}$ is as follows.
  \begin{equation*}
    \psi_{\mathcal{B}}(\mathbf{x}) 
    = 
    -\log\Prob_{\GP}(f_{\varphi}(\mathbf{x}) < 0)
    -\log\Prob_{\GP}(f_{\varphi_{12}}(\mathbf{x}) < 0)
    -\log\Prob_{\GP}(f_{\varphi_{34}}(\mathbf{x}) < 0)
  \end{equation*}
  Hence
  minimizing $\psi_{\mathcal{B}}(\mathbf{x})$ means 
  trying to falsify all the formulas $\varphi$, $\varphi_{12}$, and $\varphi_{34}$.
\end{myexample}

\begin{myremark}
  In this paper,
  we assume that 
  $\Prob_{\mathcal{B}}(\sem{\varphi} = \ffalse)$ is not 0 nor 1
  on the given Bayesian network $\mathcal{B}$.
  In the former case,
  $\psi_{\mathcal{B}}(\mathbf{x})$
  is undefined because $\Prob_{\mathcal{B}}(\sem{\varphi} = \ffalse)$ is 0,
  and the latter case,
  $\psi_{\mathcal{B}}(\mathbf{x})$ is constantly 0
  because 
  $\Prob_{\mathcal{B}}(\Theta \mid \sem{\varphi} = \ffalse) 
  = \Prob_{\mathcal{B}}(\Theta)$.
  We believe this is reasonable
  if we believe the given annotation $\mathcal{B}$ is correct---in 
  case  $\Prob_{\mathcal{B}}(\sem{\varphi} = \ffalse)$ is 0 (or 1)
  falsification never succeeds (or always succeeds, respectively).
\end{myremark}

The resulting extension of the  $\GPPS$ algorithm 
(\S{}\ref{subsec:GPOptimization}) with Bayesian networks
is presented in Algorithm~\ref{algo:GPPSatBN}.

\begin{algorithm}[htbp]
  \caption{Extension of the $\GPPS$ algorithm with Bayesian network annotation, for falsification}
  \label{algo:GPPSatBN}
  \begin{algorithmic}[1]
    \INPUT 
    an input space $\mathbf{X}$;
    a system $\mathcal{M}$;
    a specification $\varphi$ to falsity;
    a Bayesian network $\mathcal{B}$ whose nodes are labeled with formulas
    $\Phi=\{\varphi_{1},\dotsc,\varphi_{N}\}$;
    a covariance function
    $k: \mathbf{X} \times \mathbf{X} \to \R$;
    and
    an initial data set $D_{i} = \{(\mathbf{x}'_1, f_{\varphi_{i}}(\mathbf{x}'_1)), \dots, (\mathbf{x}'_s, f_{\varphi_{i}}(\mathbf{x}'_s))\}$ for each $i=1,\dotsc, N$

    \For{ $t = 1 \dots T$ }
    \State $\GP(\mu_i, k_i) 
    = \GP(\mathbf{0}, k; D_i)$ \quad for each $i=1,\dots,N$;\\
    \Comment Estimate the cost functions $f_{\varphi_{1}}, \dotsc, f_{\varphi_{N}}$ by Gaussian process regression
    \State $\mathbf{x}_{t} = \Argmin_{\mathbf{x \in X}} 
    \psi_{\mathcal{B}} (\mathbf{x})$;
    \Comment Choose a new sample input by the acquisition function
    \State $D_i = D_i \cup 
    \{(\mathbf{x}_{t}, \Robust{\mathcal{M}(\mathbf{x}_{t})}{\varphi_i})\}$
    \quad for each $i=1,\dots,N$;
    \Comment Observe the robustness
    \If{$\Robust{\mathcal{M}(\mathbf{x}_{t})}{\varphi} < 0$} \quad\Return $\mathbf{x}_{t}$;
    \Comment The specification $\varphi$ is falsified
    \EndIf
    \EndFor
  \end{algorithmic}
\end{algorithm}

\subsection{Another acquisition function based on  the difference
of KL divergence}

Aside from the acquisition function $\psi_{\mathcal{B}}$ in Def.~\ref{def:KLDist}, we propose another criteria.
\begin{mydefinition}
  [Another acquitision function $\psi'_{\mathcal{B}}(\mathbf{x})$]
  \label{def:KLDiff}
  \begin{equation*}
    \begin{split}
      \mathbf{x}
      &= \Argmin_{\mathbf{x}} \psi'_{\mathcal{B}}(\mathbf{x})  \text{ where }\\
    &
    \psi'_{\mathcal{B}}(\mathbf{x}) = 
    \KLDist\bigg(
    \Prob_{\mathcal{B}}(\Theta \mid \sem{\varphi} = \ffalse) 
    \,\bigg|\bigg|\,
    \Prob_{\GP(\mathbf{x})}(\Theta)
    \bigg)
    -
    \KLDist\bigg(
    \Prob_{\mathcal{B}}(\Theta) 
    \,\bigg|\bigg|\,
    \Prob_{\GP(\mathbf{x})}(\Theta)
    \bigg).
  \end{split}
  \end{equation*}
\end{mydefinition}

One of the advantages of this acquisition function $\psi'_{\mathcal{B}}(\mathbf{x})$ is that
we can extract it to a simpler form as follows.

\begin{align*}
  \label{eq:KLDiff}
  \psi'_{\mathcal{B}} (\mathbf{x}) =
  &
  \sum_{{\varphi_i} \in \Phi}
  \left( 
  \begin{tabular}{cll}
    & $\big(
      \Prob_{\mathcal{B}}(\sem{\varphi_i} = \ttrue) -
      \Prob_{\mathcal{B}}(\sem{\varphi_i} = \ttrue \mid \sem{\varphi} = \ffalse)
      \big)$
    & $\log$ 
    $\Prob_{\GP}(f_{\varphi_i}(\mathbf{x})  > 0)$\\
    + & $\big(
        \Prob_{\mathcal{B}}(\sem{\varphi_i} = \ffalse)
        - \Prob_{\mathcal{B}}(\sem{\varphi_i} = \ffalse \mid \sem{\varphi} = \ffalse)
        \big)$
    & $ \log $
    $\Prob_{\GP}(f_{\varphi_i}(\mathbf{x}) < 0)$\\
  \end{tabular}
\right).
\end{align*}

\begin{myexample}
 Consider the incremental counter
  in Example~\ref{ex:counter}. 
  From the Bayesian network in Fig.~\ref{fig:BNExample1Intro}
  we extract the following acquisition function
  $\psi'_{\mathcal{B}}$.
  \begin{displaymath}
    \psi'_{\mathcal{B}}(\mathbf{x})
    = \sum_{t \in [0,5]} (1 - 0.2^{t+1})
    \big(
    \log\Prob_{\GP}(f_{\varphi_t}(\mathbf{x}) > 0) - 
    \log\Prob_{\GP}(f_{\varphi_t}(\mathbf{x}) < 0)
    \big)
  \end{displaymath}
  For each formula $\varphi_t$,
  when $\Prob_{\GP}(f_{\varphi_t}(\mathbf{x}) > 0)$ becomes bigger,
  so is the value $\psi'_{\mathcal{B}}(\mathbf{x})$.
  Therefore the algorithm tries to make 
  all the formulas to be false.
  This matches our intuition in \S{}\ref{sec:Introduction}.
\end{myexample}

\begin{myexample}
 Let us consider  the automatic transmission problem in \S{}\ref{subsec:at}.
 The Bayesian network in Fig.~\ref{fig:atBayes}
  tells that 
   most of the failure of $\varphi$
  is caused by that of $\varphi_\omega$.
  The acquisition function $\psi'_{\mathcal{B}}$ is as follows.
  \begin{align*}
    \psi'_{\mathcal{B}}(\mathbf{x}) 
    & =   
      \log\Prob_{\GP}(f_{\varphi}(\mathbf{x}) > 0) - 
      \log\Prob_{\GP}(f_{\varphi}(\mathbf{x}) < 0)
      \\
    &+ 
     0.082
    \big(
    \log\Prob_{\GP}(f_{\varphi_v}(\mathbf{x}) > 0) - 
    \log\Prob_{\GP}(f_{\varphi_v}(\mathbf{x}) < 0)
    \big)\\
    &+ 
     0.817
    \big(
    \log\Prob_{\GP}(f_{\varphi_{\omega}}(\mathbf{x}) > 0) - 
    \log\Prob_{\GP}(f_{\varphi_{\omega}}(\mathbf{x}) < 0)
    \big)
  \end{align*}
  Hence as we expected,
  the satisfaction of $\varphi_\omega$ is a bigger factor than that of $\varphi_v$.
\end{myexample}

We note that extension of other (more sophisticated) acquisition functions 
(e.g. $\GPUCB$)
is not straightforward. It is one direction of our future work.

\section{Implementation and Experimental Results}\label{sec:Experiment}

\subsection{Implementation}\label{subsec:Imprementation}
Our implementation of Algorithm~\ref{algo:GPPSatBN}
consists of 
the following three open source libraries
and one new part.
They are mostly written in MATLAB.
\begin{description}
\item[Computing the robustness] We employ BREACH\cite{WEB:BreachToolbox}
  to compute the simulation output of the system $\mathcal{M}(\mathbf{x})$
  and the robustness $\Robust{\mathcal{M}(\mathbf{x})}{\varphi}$ as defined in Def.~\ref{def:semantics}.
\item[Gaussian process regression]
  Line 2 in Algorithm~\ref{algo:GPPSatBN}
  is done by GPML MATLAB Code version 4.0\cite{WEB:GPML}, 
  a widely used library for 
   computation about Gaussian processes.
\item[Inference on  Bayesian networks]
  We employ Bayes Net Toolbox for Matlab\cite{WEB:BayesNetToolbox}
  for inference on Bayesian networks.
\item[
The algorithms $\GPPS$ and $\GPPI$ aided by Bayesian networks]
  This part is new. 
   Optimization of an acquisition function $\psi$ is done
  by the following two steps: 
  1) we randomly pick  initial samples 
  $\mathbf{x}_{1}, \dots, \mathbf{x}_{100}$ 
  and compute the corresponding values of $\psi$; and 2) 
  from the minimum $\mathbf{x}_{i}$ of the one hundred,
  we further do greedy hill-climbing search.
\end{description}

\subsection{Experiments}
Using our implementation we conducted the following experiments.
We do experiments for the three falsification problems;
Problem 1 is from Examples~\ref{ex:counter},
Problem 2 from \S{}\ref{subsec:sineSystem} and
Problem 3 from \S{}\ref{subsec:at}.
For the automatic transmission example (in \S{}\ref{subsec:at}) we used two different parameters;
Problem 3-1
is with the specification 
$\varphi = \BoxOp{} (v > -1 \wedge \omega < 4780)$; and 
Problem 3-2 is with
$\varphi = \BoxOp{} (v < 120 \wedge \omega < 4780)$.

The experiments were done 
on a ThinkPad T530 with Intel Core i7-3520M
2.90GHz CPU with 3.7GB memory. The OS was Ubuntu14.04 LTS (64-bit).
A single falsification trial consists of a number of iterations---iterations 
of for-loop in line 2 in Algorithm~\ref{alg:falsificationByOptHighLevel}---before it succeeds or times out (after 100 seconds). 
For each problem
we made ten falsification trials.
We made multiple trials because of
the stochastic nature of the optimization algorithm.
We measured the performance
by the following criteria:
\begin{itemize}
  \item \textbf{Success rate:}
  The number of successful trials
   (out of ten).
\item \textbf{The number of iteration loops:}
  The average number of iteration loops
  to find the counterexample.
\item \textbf{The computational time:}
  The average time
  to find the counterexample.
\end{itemize}

Besides our two extended algorithms with 
the acquisition functions (in Def.~\ref{def:KLDist} and \ref{def:KLDiff}),
we measured the performance of 
the conventional Gaussian process optimization algorithms $\GPPS$
and compare them.

The experimental results are in Table~\ref{table:Experiment}.
We see that our causality-aided approach ($\GPPS$ with $\psi_\mathcal{B}$ and $\psi'_\mathcal{B}$) 
significantly outperformed others for Example~\ref{ex:counter}. This suggests promising potential of the proposed approach in the context of probabilistic programs---all the more because Bayesian networks like in Fig.~\ref{fig:BNExample1Intro} could be systematically derived using probabilistic predicate transformers. 

Our algorithms performed at least as well as the conventional $\GPPS$, for the other examples (Problem 2, 3-1 and 3-2). 
In Problem 3-1 and 3-2 we observe that our algorithms took fewer iterations before successful falsification. 
This is potentially an advantage when we wish to deal with  bigger Simulink models as system models $\mathcal{M}$ 
(their numerical simulation, i.e.\ computation of  $\mathcal{M}(\sigma)$, is computationally expensive). 
That said, we believe the idea of causality aid in falsification can be a breaking one, with a potential of accelerating falsification by magnitudes. 
Its current performance for Problem~3 (that is from cyber-physical systems, a main application domain of falsification) is therefore not satisfactory. 
We will therefore pursue further improvement of our algorithm (Algorithm~\ref{algo:GPPSatBN}).

\begin{table}[tbp]
\caption{Experiment results}
\label{table:Experiment}
 \scalebox{0.77}{\parbox{\linewidth}{
  \begin{tabular}{ c || c|c|c || c|c|c || c|c|c || c|c|c }
    & \multicolumn{3}{|c||}{Problem 1} & \multicolumn{3}{|c||}{Problem 2} & \multicolumn{3}{|c||}{Problem 3-1} & \multicolumn{3}{|c}{Problem 3-2}\\
    & Succ. & Iter. & Time & Succ. & Iter. & Time & Succ. & Iter. & Time & Succ. & Iter. & Time \\
    Algorithm &  & (Succ.) & (Succ.) &  & (Succ.) & (Succ.) &  & (Succ.) & (Succ.) &  & (Succ.) & (Succ.) \\\hline\hline
    $\GPPS$ & 0 & 208.2 & 100.0 & 5 & 122.4 & 77.2  & 10 & 49.5 & 41.5 & 10 & 23.0 & 6.7 \\
    &  & - & - &  & 93.6 & 54.4 &  & 49.5 & 41.5 &  & 23.0 & 6.7 \\\hline\hline
    $\GPPS$ with $\psi_{\mathcal{B}}$ & 7 & 109.3 & 81.3 & 6 & 62.5 & 77.4 & 8 & 32.0 & 56.7 & 10 & 15.7 & 7.2 \\
    &  & 105.0 & 73.2 &  & 52.7 & 62.4 &  & 28.0 & 45.8 &  & 15.7 & 7.2 \\\hline
    $\GPPS$ with $\psi'_{\mathcal{B}}$ & 5 & 104.5 & 76.3 & 5 & 63.2 & 81.0 & 7 & 36.6 & 64.3 & 10 & 13.7 & 25.0 \\
    &  & 92.0 & 52.7 &  & 51.2 & 62.1 &  & 29.8 & 49.0 &  & 13.7 & 25.0 \\
  \end{tabular}
 }}
\end{table}

\section{Future Work}
In this paper,
we show that the causality information
given in the form of a Bayesian network
helps us to solve falsification problems efficiently.
However,
we still have many challenges
in constructing such helpful Bayesian networks.
As we discussed in \S{}\ref{sec:Introduction},
we expect that 
the theory of probabilistic programming languages
will shed light on the problem,
but at any rate
we need more practical example scenarios
to evaluate the viability of our approach.

Moreover, we conceive that
our proposed algorithm in \S{}\ref{sec:Algorithm}
contains the potential for many improvements.
As we note in \S{}\ref{subsec:GPPS},
the acquisition function in $\GPPS$ is simple,
but not the state-of-the-art
in the field of Gaussian process optimization.
Extending our approach to other type of the acquisition function
is not straightforward,
but we think it is within possibility.

\bibliographystyle{eptcs} 
\bibliography{./relatedWork} %

\end{document}